\documentclass[aps,prd,nofootinbib,notitlepage, floatfix,superscriptaddress,showkeys]{revtex4-1}
\usepackage{comment} 
\usepackage{booktabs} 
\usepackage{fullpage} 
\usepackage{amsmath}
\usepackage{amssymb}
\usepackage{bm}
\usepackage{graphicx}
\usepackage{amsthm}
\usepackage[dvipsnames]{xcolor}
\usepackage[page,title]{appendix}

\DeclareMathOperator{\Tr}{Tr}

\begin{document}
\title{Stability of multi-component relativistic viscous hydrodynamics from Israel-Stewart and reproducing  DNMR from maximizing the entropy}

\author{Dekrayat Almaalol}
\affiliation{Illinois Center for Advanced Studies of the Universe, Department of Physics, University of Illinois at Urbana-Champaign, Urbana, IL 61801, USA}
\author{Travis Dore}
\affiliation{Illinois Center for Advanced Studies of the Universe, Department of Physics, University of Illinois at Urbana-Champaign, Urbana, IL 61801, USA}
\author{Jacquelyn Noronha-Hostler}
\affiliation{Illinois Center for Advanced Studies of the Universe, Department of Physics, University of Illinois at Urbana-Champaign, Urbana, IL 61801, USA}

\date{\today}
\begin{abstract}
The Quark Gluon Plasma produced in heavy-ion collisions has three relevant conserved charges: baryon number (B), strangeness (S), and electric charge (Q). Here we derive the Israel-Stewart framework for BSQ diffusion coupled to shear and bulk viscosity and the thermodynamic derivatives needed to couple this to a BSQ equation of state. We reproduce a subset of the ${\mathcal{J}}$ terms in the DNMR approach and propose a technique to derive other terms from the maximum entropy principle.  Finally, we preform a stability analysis that can be used to constraint transport coefficients in BSQ hydrodynamics.
\end{abstract}
\maketitle
%
\section{Introduction}
Within the standard model, baryon number (B) and electric charge (Q) are always conserved. 
Additionally, the time scales of relativistic heavy-ion collisions are significantly shorter than the strange quark weak decay such that strangeness (S) is also conserved. Thus, dynamical simulations of heavy-ion collisions must conserve BSQ, which is especially important at finite baryon densities $\rho_B$. Already, initial studies using only one conserved charged \cite{Denicol:2018wdp,Du:2019obx}, ideal BSQ dynamics \cite{Schafer:2021csj}, or lower dimensions \cite{Fotakis:2019nbq,Du:2021zqz} have begun to make model-to-data comparisons.  From these initial studies it was shown that the inclusion of strangeness neutrality can play a strong role \cite{Monnai:2021kgu}, fluctuations of BSQ conserved charges may affect strangeness flow \cite{Martinez:2019jbu,Carzon:2019qja}, and that BSQ diffusion affects the rapidity dependence of charge distribution \cite{Fotakis:2019nbq}. Because baryon diffusion has very non-trivial behavior at the critical point (aka baryon opalescence) wherein it drops to 0 \cite{Rougemont:2015ona,Monnai:2016kud,Rougemont:2017tlu,Grefa:2022sav}, it is that much more urgent that we understand the interplay between BSQ diffusion and how that couples to other transport coefficients (such as the bulk viscosity that diverges at the critical point \cite{Monnai:2016kud,Martinez:2019bsn,Rajagopal:2019xwg,Dore:2020jye,Dore:2022qyz}). 

At this point different approaches exist for deriving the relativistic viscous hydrodynamic equations of motion i.) Phenomenological Israel-Stewart approach wherein the equations of motion are derived directly from enforcing that the entropy current must be positive \cite{Israel:1979wp} ii.) the DNMR approach \cite{Denicol:2012cn} which derives the equations of motion from the Boltzmann
equation using the method of moments by applying a systematic power-counting scheme in Knudsen and inverse Reynolds number iii.) the BDNK equations of motion wherein a generic first-order expansion of the energy-momentum tensor and charge current is employed, leading to stable and causal equation of motion in non-traditional hydrodynamic frames \cite{Bemfica:2017wps,Bemfica:2020zjp,Kovtun:2019hdm,Hoult:2020eho}, iv.) anisotropic hydrodynamics \cite{Martinez:2010sc,Martinez:2012tu,Alqahtani:2017jwl} which implements a re-summation scheme using the moment method of kinetic theory.  It has been shown in a number of different approaches for toy model scenarios that DNMR approaches an attractor more quickly than Israel-Stewart \cite{Heinz:2015gka, Strickland:2017kux, Behtash:2017wqg}. Since BDNK is such a new theory, studies of its attracting like behavior are not as developed, although there has been some work \cite{Bemfica:2017wps}. In this work, we will only consider second-order theories. 

Because heavy-ion collisions probe the QCD phase diagram dynamically, the search for the QCD critical point depends not only on universality in the EoS but also the dynamical universality class.  Thus, the effects of critical scaling of transport coefficients may probe far-from-equilibrium dynamics close to the critical point.  In fact, it is known that the hydrodynamic picture breaks down exactly at the critical point \cite{Stephanov:2017ghc,Nahrgang:2018afz,Rajagopal:2019xwg,Dore:2020jye}, although uncertainties remain in terms of how close one can approach the critical point and still remain in a hydrodynamic regime.  While a number of advances have been made on critical scaling within hydrodynamics \cite{Stephanov:2017ghc, An:2019csj,Pradeep:2022mkf}, there is not yet one standardized approach that is agreed on within the community that is also known to preserve the properties of causality and stability in the equations of motion.  

Thus, our approach is to use hydrodynamic equations of motion that are the most stable even far-from-equilibrium.  Previously, it was  confirmed \cite{Dore:2020jye} that DNMR equations of motion appear to be the most stable (by studying attractors) when it comes to critical scaling of transport coefficients when compared to phenomenological Israel-Stewart. 
For the case of BSQ charge conservation, new relativistic viscous fluid dynamics equations must be derived taking into account the BSQ diffusion matrix \cite{Greif:2017byw}.  Initial studies have done this using a phenomenological Israel-Stewart approach in \cite{Monnai:2012jc}, and based on DNMR in \cite{Fotakis:2022usk} (although not with all possible DNMR terms).  However, there have not yet been studies that can systematically connect the Israel-Stewart approach and DNMR, nor have there been causality and stability analyses (a la \cite{Bemfica:2020xym,Plumberg:2021bme,Chiu:2021muk}) applied to BSQ diffusion yet. Here we use a new approach that can reconcile nearly all terms in the equation of motion between phenomenological Israel-Stewart and DNMR.  Then, we preform a linear stability analysis to find constraints between both the transport coefficients and the EoS. We discuss the implications of these constraints in specific limits. For instance, while the exploration of meta-stable states requires other physics beyond what is discussed here, we can at least determine if certain features of a meta-stable region like $c_s^2<0$ or a negative enthalpy are consistent with our stability analysis.  Indeed, we find that nothing prevents such features within our stability constraints.

This paper is organized as follows. In Sec. \ref{sec:EOM} we derive the Israel-Stewart-based BSQ relativistic viscous hydrodynamic equations of motion and connect them to DNMR. In Sec. \ref{sec:stability} we derive our stability constraints for BSQ hydrodynamics using two different approaches.  Finally, in Sec.\ \ref{sec:conclude} we provide our conclusions and outlook. 
\section{multi-component transient fluid dynamics with BSQ charges}
\label{sec:EOM}
Hydrodynamics is an effective field theory based on the local conservation laws of energy-momentum tensor $T^{\mu\nu}(x)$ and charge currents $N^\mu_q(x)$, 
\begin{align}
D_\mu T^{\mu\nu}= 0\,;
\qquad\qquad D_\mu N^{\mu}_q=0\,, \qquad q\in\{B,S,Q\}\,,
\label{Eq:conservation}
\end{align}
where $D_\mu$ is the covariant derivative which gives for a general coordinate system
\begin{align}
D_\mu T^{\mu\nu} = \frac{1}{\sqrt{g}}\partial_\mu (\sqrt{g} T^{\mu\nu}) + \Gamma^\nu_{\mu\alpha} T^{\mu\alpha} = 0\,,  
\end{align}
\begin{align}
D_\mu N_q^{\mu} = \frac{1}{\sqrt{g}}\partial_\mu (\sqrt{g} T^{\mu\nu}) = 0 \,, 
\end{align}
where the metric tensor $g^{\mu\nu}$ is defined with negative signature $(+,-,-,-$). By extending to multiple conserved charges, the set of equations of motion which evolve the system increases to 4+$N_q$ equations, which describe the space-time evolution of $10+4\, N_q$ independent components of the energy momentum tensor and the charges current(s), respectively.

The energy-momentum tensor and the charge currents consist of an equilibrium contribution and a dissipative part, 
\begin{align}
T^{\mu\nu}&= T^{\mu\nu}_{0} + \Pi^{\mu\nu}\,,\qquad \qquad N^{\mu}_{q}= N^{\mu}_{q,0}+ n^{\mu}_{q}\,; \qquad q\in\{B,S,Q\} 
\label{Eq:eq+noneq}
\end{align}
where the ``\,0\," subscript refer to the equilibrium contribution. The non-equilibrium contributions of the energy moment tensor is $\Pi^{\mu\nu}=\Pi\Delta^{\mu\nu}+\pi^{\mu\nu}$ where $\Pi=\frac{1}{3}\Tr \left[\Pi^{\mu\nu}\right]$ is the bulk pressure and the symmetric, traceless component of $\Pi^{\mu\nu}$ is $\pi^{\mu\nu}$, known as the shear stress tensor. The non-equilibrium contribution of the charge current for a given conserved charge $q$ is $n^\mu_q$.
The decomposition of the energy momentum tensor and charge current in the Landau frame is
\begin{align}
T^{\mu\nu} &= \varepsilon u^{\mu}u^{\nu}- (P+\Pi) \Delta^{\mu\nu} +\pi^{\mu\nu}\, ;\label{Eq:energy-Decom}\\
N^{\mu}_q &= \rho_q u^{\mu}+n^{\mu}_q\;,  \qquad \qquad q\in\{B,S,Q\}\label{Eq:charge-Decom}
\end{align}
provided the baryon (B), strange (S), and electric charge (Q) are conserved simultaneously. The time-like future directed flow vector $u^\mu$ is normalized such that $u^\mu u_\mu = 1$, and $u^\mu u^\nu$ is defined as the temporal projection operator. Similarly, $\Delta^{\mu\nu}= g^{\mu\nu}-u^\mu u^\nu$ is a spatial projection operator. The thermodynamic variables are then energy density, $\varepsilon$, the pressure, $P$, and the individual charge densities $\rho_q$.

The 4-flow vector $u^\mu$ is defined as the eigenvector for the energy-momentum tensor $u_\mu T^{\mu\nu} = \varepsilon\, T^{\mu\nu}$ with the energy density $\varepsilon$ as its eigenvalue. The energy and charge densities are hence defined in the Landau frame by 
\begin{equation}
    \varepsilon\ u^\mu = u_\nu T^{\mu\nu} \;;\qquad \rho_q = u_\mu N^{\mu}_q\;,\label{Eq:landau}
\end{equation}
where, in the local rest frame of the fluid, $u^\mu_{LRF}=(1,0,0,0)$. With this particular choice of `` hydrodynamic frame ", we have defined $\varepsilon$ as the equilibrium energy density , and $\rho_q$ as the equilibrium charge density. Eq. (\ref{Eq:landau}) is sometimes referred to as `` Landau matching " since taken together to hold generally, they imply the use of the Landau frame.  Where we then have
$u_\mu \pi^{\mu\nu} = 0; \qquad u_\mu n^\mu_q = 0\,\, \forall\,\, q\in\{B,S,Q\}$
by definition.

The best constraints on relativistic viscous hydrodynamics come from heavy-ion collisions at the LHC/top RHIC energies that corresponds to a vanishing baryon density.  Thus, Landau matching has been a natural choice in heavy-ion collisions because it eliminates extra transport coefficients and because the Eckart frame is not well defined for zero net charge.  However, in the case of diffusion, there will be now transport coefficients but one can still use the previously determined constraints in the limit of vanishing baryon densities as a starting place. 

Applying the temporal and spatial projections $u^\mu u^\nu$ and $\Delta^{\mu\nu}$ to the conservation laws in Eq. (\ref{Eq:conservation}) and using the decomposition of $T^{\mu\nu}$ and $N^\mu$ Eqs. (\ref{Eq:energy-Decom}) and (\ref{Eq:charge-Decom}) gives  
\begin{align}
 D\varepsilon &= -(\varepsilon+P)\theta - \Pi \theta + \pi_{\mu\nu} \sigma^{\mu\nu}\,;\qquad\qquad \qquad \qquad\qquad \quad \rm ``energy\,\, conservation"\label{Eq:energy_conservation}\\
(\varepsilon+P)Du^\mu +\Pi Du^\mu &= \nabla^\mu(P{+}\Pi)- \Delta^{\mu\nu} \nabla^\lambda\pi_{\nu\lambda} + \pi^{\mu\nu} Du_\nu\,;\quad \qquad \qquad  \rm ``momentum\,\,conservation"\label{Eq:momentum_conservation}\\
D \rho_q &= -\rho_q \theta -\nabla_\mu n^\mu_q\;. \qquad \qquad \qquad \qquad\qquad \qquad \qquad \quad \quad \rm ``charge \,\,conservation"
\label{Eq:charge_conservation}
\end{align}
where $D\equiv u_\mu D^\mu$ is the covariant time derivative, $\theta\equiv D_\mu u^\mu$ is the expansion scalar, and $\nabla^\mu\equiv \Delta_{\mu\nu} \partial^\nu$ is the spatial gradient in the local rest frame of the fluid. From Eq.\ (\ref{Eq:momentum_conservation}) it becomes clearer why we refer to $\Pi$ as the bulk pressure because it enters as $P+\Pi$ such that one obtains an effective pressure.

For an ideal BSQ-charged fluid Eqs. (\ref{Eq:energy_conservation} - \ref{Eq:charge_conservation}) give a set of 7 first-order PDEs for the equilibrium fields $(\varepsilon, P, u^\mu, \rho_q)$, which represent 8 independent variables; therefore, one needs an additional input to the equations. The local equilibrium assumption of hydrodynamics means that the thermodynamic equation of state, $P=P(\varepsilon, \rho_q)$, should be used to reduce the space of variables.
With that, the system is completely determined provided the initial values of  $(\varepsilon_0, P_0, u^\mu_0, \rho_{0}^q)$ are known.
\subsection{Phenomenological approach}
\label{subsec:IS}
When including viscous effects, one also needs to be able to describe the dynamical evolution of the bulk, shear stress tensor, and the diffusive BSQ charge-currents $(\Pi$, $\pi_{\mu\nu}$, $n^\mu_q)$ extending the dynamical space of variables to $(8+1+5+3 N_q)$  which requires an additional $6+3N_q$ equations for the dissipative currents to evolve the 23 - 1 independent variables where the EoS closes the equations through the relation $P= P(\varepsilon, \rho_q)$. 

Now that the equations of motions from conservation laws are established in Eqs. (\ref{Eq:energy_conservation} - \ref{Eq:charge_conservation}), we need the time evolution for the out of equilibrium contributions: $\pi^{\mu\nu}$, $\Pi$, and $n_q^{\mu}$.
We derive the dynamical equations for these currents using the phenomenological approach of extended thermodynamics originally formulated by Israel-Stewart \cite{Israel:1979wp}. In this approach, The second law of thermodynamics  
\begin{align}
 D_\mu S^\mu_{0}=  \beta_\nu D_\mu T^{\mu\nu}_{0} -  \sum_q^{B,S,Q} \alpha_q D_\mu N^\mu_{q, 0} \geq 0,
\end{align}
is extended out of equilibrium using the local conservation laws (\ref{Eq:conservation}) in the form (\ref{Eq:eq+noneq}) while constraining the entropy production by the second law of thermodynamics which then gives
\begin{align}
 D_\mu S^\mu_{0}= - \beta_\nu D_\mu T^{\mu\nu}_{non-eq} +  \sum_q^{B,S,Q} \alpha_q D_\mu N^\mu_{q, non-eq} \geq 0,
\end{align}
where $S^\mu_{non-eq}$ is the entropy current out of equilibrium. The thermodynamic quantities are defined by $\beta_\nu=\frac{u^\nu}{T}$, with T being the equilibrium temperature. The thermodynamic potential associated with the conserved charge $q$ is $\alpha_q = \frac{\mu_q}{T}$ where $\mu_q$ is the chemical potential. Finally, $T^{\mu\nu}_{non-eq}$ and $N^\mu_{q, non-eq}$ are the non-equilibrium contributions to the full energy momentum tensor and charge currents from Eq. (\ref{Eq:eq+noneq}).

The starting point is the proposition that the second-order entropy current for a multi-component fluid can be written as the following 
\begin{align}
{S}^\mu_{non-eq} &= s u^{\mu} - \sum_q^{B,S,Q} \alpha_{q}{n}^\mu_q - \frac{1}{2}u^\mu\left( \beta_\Pi{\Pi}^2 + \beta_\pi {\pi}^{\mu\nu}{\pi}_{\mu\nu}+ \sum_q^{B,S,Q} \beta_{n}^{qq'} n_q^\mu{n}_\mu^{q'} \right) +\sum_q^{B,S,Q} \left( \delta_{n\Pi}^q n^\mu_q{\Pi} + \delta_{n\pi}^{q} n^\nu_q{\pi}^{\mu}_{\nu} \right)\nonumber\\
&+ u_\nu (\delta_{\Pi \pi} \Pi \pi^{\mu\nu}) + higher\,\,order\,\, terms\,\,,
\label{Eq:entropy-current}
\end{align}
where the equilibrium entropy density, s, is given by Gibbs–Duhem equation
\begin{equation}
    \varepsilon + p = sT + \sum_q^{B,S,Q} n_q \mu_q\,, 
\end{equation}
and the equilibrium entropy current is given by $s^\mu_0=su^\mu$. The second order transport coefficients are $\{\beta_\Pi,\beta_\pi\}$ for bulk and shear stress tensor, and $\{\delta_{\Pi\pi}\}$ for coupling between ($\Pi,\pi_{\mu\nu}$). The matrix $\beta_{n}^{qq'}$ is a symmetric matrix for diffusion of charge $q\leftrightarrow q'$, and  $\{\delta_{n\Pi}^q,\delta_{n\pi}^{q}\}$ are the second order coupling coefficients between any fluid component q and bulk pressure and shear stress-tensor, respectively.
 
Extending the standard IS treatment to multiple charges by including all three charge currents $\{n_B, n_S, n_Q\}$ adds six additional independent contributions from the $\beta_{qq'}$ term which account for the diffusive coupling among these charges. It will also extend the terms $\{\delta_{n\Pi}^q,\delta_{n\pi}^{q}\}$ similarly. 

The entropy production rate at a space-time point $x$ is obtained from the four divergence of the entropy current vector in Eq. (\ref{Eq:entropy-current}), giving 
\begin{align}
    \partial_\mu{S}^\mu &= \left( -\beta \theta \Pi - \beta_\Pi\dot{{\Pi}}\Pi - \frac{\dot\beta_\Pi}{2}\Pi^2 - \frac{\beta_\Pi}{2}{\Pi^2}\theta  \right) 
    +\left( \beta \pi_{\mu\nu} \sigma^{\mu\nu} - \beta_\pi  \dot{{\pi}}^{\mu\nu}{\pi}_{\mu\nu} - \frac{\dot\beta_\pi}{2}{\pi}^{\mu\nu}{\pi}_{\mu\nu} + \frac{\beta_\pi}{2}{\pi}^{\mu\nu}{\pi}_{\mu\nu}\theta  \right)\nonumber \\
     &- \sum_q \left(  n^\mu_q \nabla_\mu\alpha_{q} + \beta_{qq'} \dot{{n}}^\mu_{q} n_\mu^{q'} + \frac{\dot{\beta}_{qq'}}{2}{n}^\mu_{q} n_\mu^{q'}+ \frac{\beta_{qq'}}{2}{n}^\mu_{q} {n}_\mu^{q'}\theta  \right)- \bigg( (\partial_\mu u_\nu)\delta_{\Pi \pi}\Pi \pi_{\mu\nu} +\delta_{\Pi\pi} \Pi (u_\nu \partial_\mu{\pi}^{\mu\nu})  \bigg)\nonumber\\
     &- \sum_q\bigg(\delta_{n\Pi}^{q} n_q^\mu \nabla_\mu{\Pi}+\delta_{n\Pi}^{q}\Pi \partial_\mu{n}^\mu_q + \lambda_{\Pi}^q{n}_q^\mu \Pi \nabla_\mu\delta_{n\Pi}^{q} -\delta_{n\pi}^{q} n^\mu_q \nabla^\nu {\pi}_{\mu\nu}+\delta^{q}_{n\pi} \pi_{\mu\nu} \nabla^{\langle\mu}n_q^{\nu\rangle} + \lambda_\pi^q \pi_{\mu\nu} n_q^{\langle\mu}\nabla^{\nu\rangle}\delta^{q}_{n\pi} \bigg)\,,
\label{eqn:div-entropy1}
\end{align}
where the subscript ``\, $\dot {}\,$\," denotes the covariant time derivative $u^\mu D_\mu$ of a particular term. The angular brackets across indices is to emphasize that after contracting with $\pi^{\mu\nu}$, only the trace-less
symmetric part survives. 

The standard treatment of Israel-Stewart is to define the tensors $\{\Pi, \pi_{\mu\nu}, n^{\mu}_q\}$ in such a way which introduces an effective description of these dissipative currents. This definition reproduces the generic structural form of second law of thermodynamics in order to guarantee the positivity of the entropy production. We follow the same procedure here and start by regrouping the dissipative terms in the entropy production to get
\begin{align}
    \partial_\mu{S}^\mu &=\beta{\Pi}\left\{ -\theta - \frac{\beta_\Pi}{\beta}\dot{{\Pi}} - \frac{\dot\beta_\Pi}{2\beta}{\Pi} - \frac{\beta_\Pi}{2\beta}{\Pi}\theta - \frac{1}{\beta}\sum_q\left(\delta_{n\Pi}^{q}\partial_\mu{n}^\mu_q + \lambda_\Pi^q{n}_q^\mu\nabla_\mu\delta_{n\Pi}^{q} \right) -  \frac{\delta_{\Pi \pi}}{2\beta} (\partial_\mu u_\nu) \pi^{\mu\nu}\right\}\nonumber\\
    & + \beta \pi_{\mu\nu}\left\{ \sigma^{\mu\nu} - \frac{\beta_\pi}{\beta}\dot{{\pi}}^{\mu\nu} - \frac{\dot\beta_\pi}{2\beta}{\pi}^{\mu\nu} + \frac{\beta_\pi}{2\beta}{\pi}^{\mu\nu}\theta - \frac{1}{\beta}\sum_q \left(\delta^{q}_{n\pi}\nabla^{\langle\mu}n_q^{\nu\rangle} + \lambda_\pi^q n_q^{\langle\mu}\nabla^{\nu\rangle}\delta^{q}_1) \right)- \frac{\delta_{\Pi \pi}}{2\beta} (\partial_\mu u_\nu) \Pi \right\}\nonumber\\
    &+ \sum_q {n}^\mu_q\sum_{q'}\left\{ -\nabla_\mu\alpha_{q'} - \beta_{qq'}\dot{{n}}_\mu^{q'} - \frac{\dot{\beta}_{qq'}}{2}{n}_\mu^{q'} - \frac{\beta_{qq'}}{2}{n}_\mu^{q'}\theta- \left(\delta_{n\Pi}^{q}\nabla_\mu{\Pi} -(1-\lambda_\Pi^q){\Pi}\nabla_\mu\delta_{n\Pi}^{q} \right) 
    - \delta_{n\pi}^{q}\nabla^\nu{\pi}_{\mu\nu} \right.\nonumber\\
    &- (1-\lambda_\pi^q){\pi}_{\mu\nu}\nabla^\nu\delta_{n\pi}^{q}  \left.\right\} \,,
\label{eq:div-entropy2}
\end{align}
where while grouping we redistributed the terms  $\lambda_\Pi^q{n^\mu_q\Pi}\nabla_\mu\delta_{n\Pi}^{q}$ and $(1-\lambda_\Pi^q){n^\mu_q\Pi}\nabla_\mu\delta_{n\Pi}^{q}$ between the bulk and the charge currents relaxation equations are to account for the arbitrary contribution of the second order coupling terms $n^\mu_q \Pi$, and similarly for $n^\mu_q \pi_{\mu\nu}$. 

%
%

Both $\lambda_\Pi^q,\lambda_\pi^q$ can vary in the range between $\{0,1\}$ to set their relevant weight. Note the the coupling terms $n^\mu_q \Pi$ and $n^\mu_q \pi_{\mu\nu}$ have been ignored previously for studies of baryon diffusion e.g . \cite{Denicol:2018wdp}; however, there has been no check made to determine their role or significance. We include these terms here for their relevance especially when diffusion exists for multiple charges.

Imposing the second law of thermodynamics guarantee positive definiteness of the entropy production which, therefore, must be quadratic in dissipative currents leading to a non-negative divergence of the entropy current $S^\mu$. This requires making everything in "big brackets" in Eq. (\ref{eq:div-entropy2}) equal to its corresponding dissipative current such that
\begin{align}
\nabla_\mu S^\mu= \frac{ \Pi \Pi}{\zeta T} + \frac{ \pi^{\mu\nu} \pi_{\mu\nu}}{2\eta T}+
\frac{n^\mu_q n_\mu^{q'}}{\kappa_{qq'}T^2}\, \geq 0\,
\label{Eq:2nd-law}
\end{align}
Thus, to ensure positive entropy produce, we define
\begin{align}
\Pi &= \zeta \left\{ -\theta - \frac{\beta_\Pi}{\beta}\dot{{\Pi}} - \frac{\dot\beta_\Pi}{2\beta}{\Pi} - \frac{\beta_\Pi}{2\beta}{\Pi}\theta - \frac{1}{\beta}\sum_q\left(\delta_{n\Pi}^{q}\partial_\mu{n}^\mu_q + \lambda_\Pi^q{n}_q^\mu\nabla_\mu\delta_{n\Pi}^{q} \right)-  \frac{\delta_{\Pi \pi}}{2\beta} \pi_{\mu\nu} \sigma_{\mu\nu}\right\}\,,
\label{IS-bulk}
\end{align}
\begin{align}
    {\pi}^{\mu\nu} &= 2\eta\left\{\sigma^{\mu\nu} - \frac{\beta_\pi}{\beta}\dot{{\pi}}^{\mu\nu} - \frac{\dot\beta_\pi}{2\beta}{\pi}^{\mu\nu} + \frac{\beta_\pi}{2\beta}{\pi}^{\mu\nu}\theta - \frac{1}{\beta}\sum_q \left(\delta^{q}_{n\pi}\nabla^{\langle\mu}n_q^{\nu\rangle} + \lambda_\pi^q n_q^{\langle\mu}\nabla^{\nu\rangle}\delta^{q}_{n\pi}) \right)- \frac{\delta_{\Pi \pi}}{2\beta} \Pi  \sigma_{\mu\nu}\right\} \,,    
    \label{IS-shear}
\end{align}
\begin{align}
        {n}_\mu^q &= \sum_{q'}\kappa_{qq'}\left\{-\nabla_\mu\alpha_{q'} - \beta_{lq'}\dot{{n}}_\mu^{l} - \frac{\dot{\beta}_{lq'}}{2}{n}_\mu^{l} - \frac{\beta_{lq'}}{2}{n}_\mu^{l}\theta - \delta_{n\Pi}^{q'}\nabla_\mu{\Pi} - \delta_{n\pi}^{q'}\nabla^\nu{\pi}_{\mu\nu}-\Tilde{\lambda}_\Pi^{q'}{\Pi}\nabla_\mu\delta_0^{q'}- \Tilde{\lambda}_\pi^{q'}{\pi}_{\mu\nu}\nabla^\nu\delta_1^{q'} \right\}\,,
\label{def-IS-currents}
\end{align}
By construction, Eqs.\ (\ref{IS-bulk}-\ref{def-IS-currents}) should be constrained to reduce to the Navier-Stokes (NS) equations when the higher order terms in the entropy vanish leading to
\begin{align}
\Pi^{NS}\equiv-\zeta\theta\,, \qquad\qquad \pi_{\mu\nu}^{NS}\equiv2\eta\sigma_{\mu\nu}\,,\qquad\qquad \left(n^\mu_q\right) ^{NS}\equiv \kappa_{qq'} \nabla^\mu\alpha_{q'},    
\end{align}
which leads to the definition of the relaxation times for the shear, bulk, and diffusion and transport coefficients,
\begin{align}
\tau_\pi = 2\beta_\pi\eta/\beta, \,\qquad\qquad\tau_\Pi = \beta_\Pi\zeta/\beta, \qquad\qquad \,\, \tau_{qq'} = \beta_{ql} \kappa_{q'l}/\beta\,\label{Eq:rel-time}    
\end{align}
where now both $\tau_i$ and the transport coefficients go to zero as the second order terms vanish and the system approaches its local equilibrium state. Note that $\tau_{qq'}$ is now a $3\times3$ matrix defined via Eq. (\ref{Eq:rel-time}). We had made the assumption that $\kappa_{qq'}$ is an invertible matrix in Eq.(\ref{Eq:2nd-law}) which combined with the condition that $\beta_{qq'}$ is also an invertible matrix defines the relaxation time matrix for diffusion to be invertible as well. Note that, physically, all of these matrices should be real and symmetric and therefore the assumption of invertability makes sense.

Finally, we obtain the equations of motion for the bulk pressure $\Pi$,
\begin{align}
    \tau_\Pi \dot\Pi + \Pi& = -(\zeta + \frac{\tau_\Pi}{2}\Pi)\theta - \frac{\tau_\Pi}{2\beta_\Pi}\dot\beta_\Pi\Pi
    - \frac{\zeta\delta_{n\Pi}^{q} }{\beta}\partial_\mu n^\mu_q + \frac{\zeta \lambda_\Pi^q}{\beta}n_q^\mu\nabla_\mu\delta_{n\Pi}^{q}-  \frac{\zeta \delta_{\Pi \pi}}{2\beta} \pi^{\mu\nu} \sigma_{\mu\nu}\,,
    \label{Eq:bulk-evol}
\end{align}
the shear stress tensor $\pi^{\mu\nu}$,
\begin{align}
    \tau_\pi \dot\pi^{\mu\nu} + \pi^{\mu\nu}& = 2\eta\sigma^{\mu\nu} + \frac{\tau_\pi}{2}\pi^{\mu\nu}\theta- \frac{\tau_\pi\dot\beta_\pi}{2\beta_\pi}\pi^{\mu\nu}
    - \frac{2\eta\delta^{q}_{n\pi}}{\beta}  \nabla^{\langle\mu}n_q^{\nu\rangle} + \frac{2\eta \lambda_\pi^q}{\beta}n_q^{\langle\mu}\nabla^{\nu\rangle}\delta^{q}_{n\pi}-  \frac{\eta \delta_{\Pi \pi}}{\beta} \Pi \sigma_{\mu\nu}\,,
     \label{Eq:shear-evol}
\end{align}
and the BSQ diffusive charge currents $n^\mu_q$,
\begin{align}
    \tau_{qq'}\dot n^\mu_{q'} + n^\mu_q & = -\kappa_{qq'}\nabla^\mu\alpha_{q'} + \frac{\tau_{qq'}n^\mu_{q'}}{2\beta}\theta - \frac{\tau_{qq'}}{2\beta}\dot\beta_{q'l}n^\mu_{l}
    - \frac{\kappa_{qq'}\delta_{n\Pi}^{q'}}{\beta} \nabla^\mu\Pi- \frac{\kappa_{qq'}\delta_{n\pi}^{q'}}{\beta} \nabla_\nu\pi^{\mu\nu}\nonumber\\
    &- \frac{\kappa_{qq'}\Tilde{\lambda}_\Pi^{q'}}{\beta}\Pi\nabla^\mu\delta_{n\Pi}^{q'}- \frac{\kappa_{qq'} \Tilde{\lambda}_\pi^{q'}}{\beta}\pi^{\mu\nu}\nabla_\nu\delta_{n\pi}^{q'}\,. 
    \label{Eq:diffusion-evol}
\end{align}
where $\{\tau_\Pi,\tau_\pi,\tau_{qq'}\}$ are the microscopic time scale of the dissipative fluxes; therefore, Eqs. (\ref{Eq:bulk-evol}-\ref{Eq:diffusion-evol}) are relaxation type equations of the viscous and diffusive currents, where basically, the effective fluid description of the system emerges as the these currents relax towards their equilibrium counterparts. 

These equations, in addition to the conservation laws  Eqs.\ (\ref{Eq:energy_conservation}-\ref{Eq:charge_conservation}) provide the space-time evolution for the transient Israel-Stewart theory  multi-component fluid with BSQ charges. An addition of the formalism we introduced here is introduction of the $\pi_{\mu\nu}\Pi$ term which we discuss further in Sec. (\ref{subsec:DNMRconnection}). In this formalism, we focus on the proper treatment of the second order coupling terms between shear-bulk, shear-diffusion, and bulk-diffusion. This will allow us to systematically scrutinize and understand the interplay between these dynamics.
Finally, having included the full BSQ charges mixes the time derivatives of the charge currents through the term $\tau_{qq'} \dot{n}^\mu_{q'}$ which renders these dynamics sensitive to the chemical content of the fluid. 
 \subsection{Thermodynamics input: Equation of State}
 \label{sec:thermo} 
The equations of motion up to this point are generic for any charged fluid and carry no information on the type of matter (beyond the number of conserved charges within the system) nor the type of microscopic interactions the constituents of the system will have in order to evolve back to equilibrium. The relevant information for the thermodynamics of the QCD matter and the transport properties are provided through the Equation of State (EoS) and transport coefficients, respectively. Ideally, the transport coefficients would be linked directly to the EoS such that  one could calculate the 4D EoS directly from lattice QCD and all corresponding transport coefficients.  The reality is that not all  EoS can directly provide the transport coefficients (for instance, the 4D lattice QCD reconstructed EoS is only a Taylor series \cite{Noronha-Hostler:2019ayj,Monnai:2019hkn} and does not provide any information about transport coefficients). 
 
For 3 conserved charges, one requires a 4-dimensional (4D) EoS since one then must include temperature and BSQ chemical potentials $\left(T,\mu_B,\mu_S,\mu_Q\right)$.  For most EoS models, it is more natural to calculate thermodynamic observables along grids of $\left(T,\mu_B,\mu_S,\mu_Q\right)$.  Thus, it is also significantly easier to take derivatives along trajectories across $\left(T,\mu_B,\mu_S,\mu_Q\right)$, rather than along lines of constant entropy, $s$, for instance. However, relativistic hydrodynamics does not directly obtain $\left(T,\mu_B,\mu_S,\mu_Q\right)$, instead it is more natural to use energy density $\varepsilon$ or entropy density $s$ combined with the 3 BSQ densities i.e. $\left(\varepsilon,n_B,n_S,n_Q\right)$ or $\left(s,n_B,n_S,n_Q\right)$. Thus, there is always a mapping issue wherein one must translate from the more natural variables of the EoS $\left(T,\mu_B,\mu_S,\mu_Q\right)$ into the more natural variables for hydrodynamics $\left(s,n_B,n_S,n_Q\right)$.  
\subsection{Transport coefficients}
\label{subsec:transport}
Starting with the bulk $\Pi$ equations of motion Eq. (\ref{Eq:bulk-evol}) , we have 10 independent transport coefficients for 3 conserved charges (note we count $\tau_\Pi$ and $\zeta$ separately even though they can be related because there is a coefficient that depends on the microscopic framework). If we have X number of conserved charges, then the number of transport coefficients for $\Pi$ is
$
    N^{\Pi}_{trans\;coef.}=4+2\cdot X
$
such that for $X=3$, we obtain $N^{\Pi}_{trans\;coef.}=10$.
For the shear stress tensor $\pi^{\mu\nu}$ relaxation equation Eq. (\ref{Eq:shear-evol}), we again have 10 transport coefficients, however, one also appears in the $\Pi$ equation of motion so we only have 9 new independent transport coefficients.  We can then write the formula for the number of $\pi^{\mu\nu}$ transport coefficients as
$
    N^{\pi^{\mu\nu}}_{trans\;coef.}=4+2\cdot X$ and if $
    \textrm{$\zeta>0$ then}\;  N^{\pi^{\mu\nu}}_{trans\;coef.} = N^{\pi^{\mu\nu}}_{trans\;coef.}-1
$
such that if the bulk viscosity is nonzero then we subtract one transport coefficient because $\delta_{\pi\Pi}$ appears in both equations of motion.
Finally, for the diffusion equations \ref{Eq:diffusion-evol}, if we have X number of conserved charges, we gain new diffusion transport coefficients according to 
$
    N^{diff}_{trans\;coef.}=2\cdot \left(\sum_{i=1}^X i\right)+X\cdot 4 
$
such that for $X=3$ conserved charges we end up 24 tranport coefficients.  Our final number of transport coefficients is then
\begin{equation}
     N_{trans\;coef.}= N^{\Pi}_{trans\;coef.}+ \left(N^{\pi^{\mu\nu}}_{trans\;coef.}-1\right)+ N^{diff}_{trans\;coef.}
\end{equation}
where for the BSQ conserved charges we have 10+9+24=43 independent transport coefficients.

The transport coefficients $\{\beta_i,\tau_i,\delta_i\} \,{\forall} \,i\in \{\Pi, \pi^{\mu\nu},n^\mu\}$ are dynamical quantities and it is not possible to know the functional form of an associated transport coefficients based on the Israel-Stewart theory. Instead, we require input from a microscopic theory. Calculations of transport coefficients has been done in a weakly-coupled approach from kinetic theory \cite{Denicol:2012es,Denicol:2012cn,Molnar:2013lta,Fotakis:2019nbq,Greif:2017byw,Rose:2020sjv}, hadron resonance gas \cite{NoronhaHostler:2008ju,Pal:2010es,Khvorostukhin:2010aj,Tawfik:2010mb,Alba:2015iva,Ratti:2010kj,Tiwari:2011km,NoronhaHostler:2012ug,Kadam:2014cua,Kadam:2015xsa,Kadam:2015fza,Kadam:2018hdo,Mohapatra:2019mcl}, or transport theory \cite{Wesp:2011yy,Ozvenchuk:2012kh,Rose:2017bjz,Rais:2019chb} as well strongly coupled approach from holography \cite{Kovtun:2004de,Casalderrey-Solana:2011dxg,Finazzo:2013efa,Finazzo:2014cna,Rougemont:2015ona,Rougemont:2017tlu,Grefa:2022sav}. Current holographic calculations include one conserved charge such as  baryon conservation but can at least consider small perturbations in the strangeness or electric charge sector. On the kinetic theory side, recent effort have been done to understand and calculate the dependence of the second order transport coefficients on chemical potential. The final form for the relaxation times will rely on the approach we take to extract the transport coefficients which in turn will directly affect the dynamics and trajectories. Here we list the functional form of the transport coefficients and arrange the proper matching with our equations.

In the case of $\mu_q=0$, the shear and bulk relaxation times are provided by \cite{Denicol:2014vaa} 
\begin{align}
\frac{\zeta}{\tau_\Pi}&=15\left(\frac{1}{3}-c^2_s\right)^{2}(\varepsilon + p)\,,\qquad\qquad\qquad \frac{\eta}{\tau_\pi}=\frac{\varepsilon+p}{5}\,, \label{betaPi}
\end{align}
And for the second order transport coefficients
\begin{align}
\frac{\delta_{\Pi\Pi}}{\tau_\Pi}=\frac{2}{3};\qquad\qquad\qquad\frac{\delta_{\pi\pi}}{\tau_\pi}&=\frac{4}{3}\,;\qquad\qquad\qquad\quad\frac{\lambda_{\pi\Pi}}{\tau_\pi}=\frac{6}{5};\\
\frac{\tau_{\pi\pi}}{\tau_\pi}&=\frac{10}{7};\qquad\qquad\quad\quad\frac{\lambda_{\Pi\pi}}{\tau_\Pi}=\frac{8}{5}\left(\frac{1}{3}-c^2_s\right)\,.\nonumber
\end{align}
For a one component massless gas in the Boltzmann limit \cite{Denicol:2018wdp}
\begin{align}
\frac{\kappa_B}{\tau_{B}} &=  \,n_B \left( \frac{1}{3}\coth \left(\alpha_B \right) - \frac{n_B T}{\varepsilon + p} \right) .
\label{eq:BB}
\end{align}
where at this point, only diagonal terms of the diffusion matrix available from kinetic theory calculations \cite{Fotakis:2022usk}.
A particular feature of the Israel-Stewart formalism is the inclusion of time and spatial derivative of thermodynamic potentials $\{\alpha_q\}$ as well as derivatives of the second order transport coefficients $\{\beta_\Pi, \beta_\pi,\beta_{qq'}, \delta_0^q,\delta_1^q\}$. These transport coefficients turns out to be complex combinations of thermodynamics quantities for the case of finite charges. While these terms have been ignored mostly in the literature, \cite{Dore:2020jye} showed that they could impact the trajectories of the dynamical evolution. This might end up an important effect when it comes to the search for critical point; therefore, we take a step forward and calculate these thermodynamic derivative quantities. To do that an additional thermodynamics input from the equation of state is needed, but in principle calculate therms like ${\dot{\beta_\Pi}}$ with the aid of Eq. (\ref{Eq:energy_conservation}). Doing so we can see that it incorporates some higher order terms beyond second order in dissipative currents. In addition, these terms could be in particular important at the transition region, for instance, at the QCD critical point it is expected to have $c_s^2\rightarrow 0$ (e.g. see \cite{Parotto:2018pwx,Karthein:2021nxe}). Since the $\zeta$ scales with $\left(\frac{1}{3}-c^2_s\right)^{2}$ then a dip in $c_s^2$ leads to a peak in $\zeta $. As we will see in Sec. \ref{sec:stability}, our choice of the underlying microscopic interactions will definitely impact the regime of applicability of these equations through their dependence and connection to the EoS. Whether the analysis will be biased to a particular set or not is something we will explore in the future.
\subsection{Connections with DNMR}
\label{subsec:DNMRconnection}
To make connections to the DNMR framework \cite{Denicol:2012cn}, we start by writing the general form of the relaxation equations as a systematic expansion in Knudsen and inverse Reynolds numbers such as
\begin{align}
\dot{Re}_i^{-1} + Re_i^{-1} &= \textrm{Navier-Stokes} + {\mathcal{J}}+ {\mathcal{K}}+ {\mathcal{R}}\,,    
\end{align}
where $Re_i, Kn_i$ are the Reynolds and Knudsen numbers for the particular dissipative current i$\in \{\Pi, \pi^{\mu\nu},n^\mu_q\}$. The ${\mathcal{J}}$, ${\mathcal{K}}$, and ${\mathcal{R}}$ terms are proportional to  $Kn_i \,Re_i^{-1}$,  $Kn_i Kn_j$, and  $Re_i^{-1} Re_j^{-1}$, respectively.
The general form of the relaxation-type equations in IS theories has particular features which differ from DNMR, and we would like to discuss some of them in this section. 
 
Consider for example the bulk relaxation equation Eq. (\ref{Eq:bulk-evol}),
\begin{align}
    \tau_\Pi \dot\Pi + \Pi& = -(\zeta + \frac{\tau_\Pi}{2}\Pi)\theta - \frac{\tau_\Pi}{2\beta_\Pi}\dot\beta_\Pi\Pi
    - \frac{\zeta\delta_{n\Pi}^{q} }{\beta}\partial_\mu n^\mu_q + \frac{\zeta \lambda_\Pi^q}{\beta}n_q^\mu\nabla_\mu\delta_{n\Pi}^{q}-  \frac{\zeta \delta_{\Pi \pi}}{2\beta} \pi^{\mu\nu} \sigma_{\mu\nu}\,,
\end{align}
For convenience, we write the equation in the following form,
\begin{align}
   \tau_\Pi \dot{Re}^{-1}_\Pi + Re^{-1}_\Pi &=  K_\Pi  -  \frac{\tau_\Pi \dot{\beta_\Pi}}{2 \beta_\Pi} Re^{-1}_\Pi + \frac{\tau_\Pi}{2} K_\Pi Re^{-1}_\Pi + \frac{\zeta \lambda_{\Pi}^q \nabla_\mu \delta_{n\Pi}^q}{\beta} Re^{-1}_{n_q}- \frac{\zeta \delta_{n\Pi}}{\beta}  \partial_\mu{Re}^{-1}_{n_q}- \frac{\zeta\delta_{\Pi\pi}}{2\beta} K_\pi Re^{-1}_\pi\,,
   \label{Eq:rel-bulk-expansion}
\end{align}
 with $K_\Pi\equiv \frac{\tau_\Pi}{\theta}$ and $Re_\Pi^{-1} \propto \Pi$.  It has been well known previously that IS theories fail to reproduce the systematic expansion of DNMR theories. DNMR theories on the other hand does not have those terms with time derivatives of thermodynamics potentials $\{\dot{\beta}_\Pi, \dot{\beta}_\pi, \dot{\beta}_{qq'}\}$. 
 
 In addition, the DNMR formalism implements an irreducible set of basis to build the moments of the distribution function out of equilibrium.  This guarantees that dissipative fluxes created from the moments of the Boltzmann equation are independent and orthogonal set of dynamical degrees of freedom. This feature has not been enforced in IS theories where no symmetry arguments was imposed on the dissipative fields.
 
 The authors of DNMR have addressed in \cite{Denicol:2012cn, Molnar:2013lta} the relative importance of the above ${\mathcal{J}}$, ${\mathcal{K}}$, and ${\mathcal{R}}$ terms in their framework. The ${\mathcal{K}}$ terms were shown to render the equations of motion non-hyperbolic leading to instability of the evolution, and this issue would be recovered only by including higher order terms. The ${\mathcal{R}}$ terms on the other hand arise from the non-linearities of the collision kernel in the Boltzmann equation which signals their potential relevance in the far from equilibrium regime; otherwise, it was shown in  \cite {Molnar:2013lta} that, in the Boltzmann limit of mass-less gas with a constant cross section, these contribution can be ignored. This leaves us with the ${\mathcal{J}}$ $\propto$ $Kn_i \,Re_j^{-1}$ which has been used in most codes that implement DNMR theory.  
Interestingly, in the Israel-Stewart approach, one can actually still reproduce at least a subset of the DNMR terms by including contributions to the entropy production which has a zero contribution to the entropy current itself. An example is the term $\propto \pi_{\mu\nu} \sigma^{\mu\nu}$ from Eq. (\ref{Eq:rel-bulk-expansion}) which does not exist in the standard IS formalism. It is not clear to us why these important terms in DNMR comes up from zero net-entropy current terms in the IS formalism. 

One interpretation of the shear-bulk coupling term in the equations of motion is that it originates from a thermodynamically reversible process i.e. this process should be symmetric under reversing the flow velocity. For this reason, one would just associate a zero transport coefficient for this term at the level of the entropy current.  Provided the above argument, one could look at the level of entropy production where the 2nd law does not uniquely specify the definition of entropy production and it only constrain it to be positive definite. In this sense, we have some freedom in adding and subtracting terms as long as we respect the order of the entropy current.
Other zero entropy terms can be treated on the same footing to generate the remaining ${\mathcal{J}}$ terms. For the term $\propto \pi_{\mu\nu} \sigma^{\mu\nu}$, it was interesting to see that it is actually the original first order NS term where as $\tau_\pi \rightarrow \infty$ and as the shear stress approaches it NS limit $\pi_{\mu\nu} = 2\eta \sigma_{\mu\nu} $ the above contribution becomes $\propto \sigma_{\mu\nu} \sigma^{\mu\nu}$.
\section{Stability analysis of the BSQ hydrodynamics}
\label{sec:stability}
A critical benchmark for any hydrodynamics theory is to pass the stability test, which warrant the existence of a steady stable equilibrium state. Any `` small " perturbations around this equilibrium background should decay quickly and the system should asymptotically converge back to its equilibrium state. Growth of perturbations, on the other hand, signals an instability of the equilibrium state which then means that the configuration space cannot be physically bounded. For Israel-Stewart theory, a linear stability analysis has been done in both Landau and Eckart frames for the case of a single charge  \cite{Hiscock:1983zz, Olson:1990rzl}. In these works, the authors showed that in the linear regime Israel-Stewart theories are causal if they are stable. In a more recent work, Gavassino et al \cite{Gavassino:2021kjm}, further studied the relationship between causality and stability and demonstrated that while causality alone does not guarantee stability, a theory can be proven to be causal if it is thermodynamically stable.

In this section, we follow the method developed in \cite{Gavassino:2021cli} to (1) derive the Lyapunov functional for the entropy current in the Landau frame and generalize it to a fluid with multiple charges, and (2) systematically extract the conditions for the positivity of the energy functional. We finally proceed with the potential importance of the results by (a) making connections with another fundamental concept of the physical system which is the causality of signal propagation in this theory, and (b) check if the analysis is biased to a particular choice of transport coefficients. The entropy current in Eq. (\ref{Eq:entropy-current}) accounts for all dissipative tensors up to second order as well as all possible coupling terms with non-zero entropy production. The total entropy of a space-like hyper-surface $\Sigma$ is given by 
\begin{align}
S[\Sigma]&= \int_{\Sigma} {S}^\mu d\Sigma_{\mu},
\label{eqn:totentropy}
\end{align}
where for a perfect fluid this entropy current is conserved; however, dissipation processes and diffusion lead to an increase in the entropy. For an isolated system, and starting from a non-equilibrium state, the total entropy is a Lyapounov function and must be positive definite function of time. If we evaluate Eq. (\ref{eqn:totentropy}) on any later surface $\Sigma'$ its value must increase
\begin{align}
S(\Sigma')- S(\Sigma)&= \int_{\Sigma} \nabla_{\mu}{S}^\mu dV\,.
\end{align}
which means as the system evolve, the entropy increases until it reaches its maximum value. This maximum is unique and is achieved only for the global thermal equilibrium state. In this sense, the equilibrium state is an attractor for all thermodynamically consistent configurations of the non-equilibrium states.
If we now `` linearly " perturb the global thermodynamic state with a small perturbation, we can quantify the entropy difference between the perturbed and non-perturbed state as 
\begin{align}
\Phi_E \equiv S(\Sigma)- \Tilde{S}(\Sigma)&= \int_{\Sigma} {E}^\mu d\Sigma_{\mu}\,,
\end{align}
where the ``$\,\,\Tilde{}\,\,$" refers to the entropy of the perturbed state, $\Phi_E$ is the flux of the current $E^\mu$ which carries information on the magnitude and propagation of perturbations. Our goal then is to quantify the conditions that bound the fluctuations from growing indefinitely and also to drive the system back to its equilibrium state as  $\,t\rightarrow \infty$.

We start  by first linearly perturbing our hydrodynamical fields about the state of equilibrium up to second order in fluctuations. Then, defining the equilibrium state by setting all disspative currents to zero, see App. \ref{app:stability}, gives the following 
\begin{align}
\delta S^\mu_{non-eq}&= \left(\varepsilon + p - \mu_q \,\rho_q \right)\delta u^{\mu} + \left(\delta \varepsilon - \mu_q \,\delta \rho_q- \delta \mu_q \,\delta \rho_q + \delta s\delta T\right) u^{\mu}+ \left(\delta \varepsilon - \mu_q \,\delta \rho_q\right)\delta u^{\mu}\nonumber\\
&-\frac{\mu_q}{T}\delta n_q^{\mu} - \frac{1}{2}u^\mu \left( \beta_\Pi\delta \Pi\delta \Pi+ \beta_\pi\delta {\pi}^{\mu\nu}\delta {\pi}_{\mu\nu}+ \sum_q\beta_{n}^{qq'}\delta n_q^\mu\delta {n}_\mu^{q'}\right)\nonumber\\
&- \sum_q \left(\delta_{n\Pi}^q\delta n^\mu\delta \Pi +\delta_{n\pi}^{q}\delta n_\nu^q\delta \pi^{\mu\nu}\right)\,,
\label{Eq:tot-entropy-pert}
\end{align}
here the $\delta$ refer to the perturbed field. To drive the system to its equilibrium state, these perturbations must strictly conserve both energy-momentum and the BSQ charge currents during the full evolution. Therefore, we impose a sort of `` landau matching" on the fluctuations. Starting from the perturbed form of the landau condition $u_{\mu} T^{\mu\nu}= \varepsilon u^{\nu}$ one gets
\begin{align}
u_{\nu}\delta T^{\mu\nu}&=  \varepsilon\delta u^{\mu} +\delta \varepsilon u^{\mu} +\delta \varepsilon\delta u^{\mu}-\delta u_{\nu} T^{\mu\nu} -\delta u_{\nu}\delta T^{\mu\nu}\,,
\end{align}
which provides the zero energy flux contributions to the entropy fluctuations. Likewise, the charge current $N^{\mu}_q= \rho  u^{\mu} + n^{\mu}_q$ can be expanded in perturbed form to give
\begin{align}
\delta N^{\mu}_q&=  \rho_q \,\delta u_{\mu} +\delta \rho_q\, u_{\mu} +\delta \rho_q \,\delta u_{\mu} +\delta n^{\mu}_q \,,
\end{align}
which provides the zero charge flux contributions to the entropy fluctuations. Note that this condition will impose conditions on each of the charges separately
Subtracting out the zero flux contributions using the killing vector and conservation laws Eq. (\ref{Eq:landau}), we can  finally define the information current $E^\mu$ such that  
\begin{align}
E^{\mu}&= \delta s^\mu - \textrm{zero energy flux}- \textrm{zero number flux  }\,,  
\end{align}
where now using the perturbed entropy form Eq. (\ref{Eq:tot-entropy-pert}) then gives the final form of the energy functional
\begin{align}
E^{\mu}&=- (\varepsilon+ {p})\delta u^{\nu}\delta u_{\nu} u^{\mu} +\delta {p}\,\delta u^{\mu} + \frac{u^{\mu}}{2 (\varepsilon + {p})}  \left. \frac{\partial \varepsilon}{\partial {p}}\right\vert_{s} (\delta{p})^2 +  \frac{u^{\mu}}{2 (\varepsilon + {p})}\left.\frac{\partial \varepsilon}{\partial s}\right\vert_p \left.\frac{\partial {p}}{\partial s}\right\vert_{\alpha_q} (\delta s)^2 +  \left. \frac{\partial\alpha_q}{\partial {p}}\right\vert_s\delta {p}\,\delta n_q^{\mu}\nonumber\\
&+ \left. \frac{\partial}{\partial\alpha_q}\right\vert_{p}\delta s \,\delta n_q^{\mu} +\delta\Pi \,\delta u^{\mu}+\delta \pi^{\mu\nu}\delta u_{\nu}+ \frac{1}{2}u^\mu \left( \beta_\Pi\delta \Pi\delta \Pi+ \beta_\pi\delta {\pi}^{\mu\nu}\delta {\pi}_{\mu\nu}+ \sum_q\beta_{n}^{qq'}\delta n_q^\mu\delta {n}_\mu^{q'}\right)\nonumber\\
&+ \sum_q \left(\delta_{n\Pi}^q\delta n^\mu\delta \Pi +\delta_{n\pi}^{q}\delta n_\nu^q\delta \pi^{\mu\nu}\right)\,.
\label{eq:Lyabunov}
\end{align}
The information current $E^\mu$ carries all information about the perturbation and can be written as 
\begin{align}
    \nabla_\mu E^\mu= -\left[\frac{\delta \Pi \delta \Pi}{\zeta T} + \frac{\delta \pi^{\mu\nu} \delta \pi_{\mu\nu}}{2\eta T}+
    \frac{\delta n^\mu_q \delta n_\mu^q}{\kappa_{qq'}T^2}\right]\,,
    \label{Eq:E}
\end{align}
where now one can state that
\begin{align}
\nabla_\mu E^\mu \leq 0; \qquad \iff \rm provided \, if \,and\, only \,if \qquad \nabla_\mu S^\mu \geq 0\,,
\label{Eqn:divE}
\end{align}
which is a direct consequence of the second law of thermodynamics. The relation Eq. (\ref{Eqn:divE}) is therefore interpreted as the loss of information about the perturbations as the system evolves towards its equilibrium state which minimizes its energy and maximizes its entropy. 
The total energy associated with $E^{\mu}$ can then be found by integration of the above expression over the space-like surface $\Sigma$
\begin{align}
E(\Sigma)&= \int_{\Sigma} E^{\mu} d\Sigma_{\mu} \qquad= \int e \beta^{\mu} dx_{\mu} \nonumber\,,
\end{align}
here $\beta^\mu=u^\mu/T$ and e is defined as the energy density of $E^\mu$ given by $e=   \frac{TE^{\mu}n_{\mu}}{u^{\nu} n_{\nu}} $. The vector $n_{\mu}$ is the future directed unit normal to $\Sigma$. Since $E$ can not increase in time, we have that 
\begin{align}
E^\mu n_\mu \geq 0\,,
\end{align}
The stability of the theory is now contingent on the proof that the energy density of the energy current Eq. (\ref{eq:Lyabunov}) is  positive definite. 
\subsection{Stability conditions}
\label{subsec:conditions}
We include the detailed method of derivation and the steps for extracting the conditions in App. \ref{app:stability}. After contracting the energy current $E^\mu$ with the future directed time-like vector $n_\mu$, the energy density is given by
\begin{align}
e&=- (\varepsilon+ {p})\delta u^{\nu}\delta u_{\nu}  +\delta {p}\,\delta u^{\mu} \lambda_\mu + \frac{1}{2 (\varepsilon + {p})}  \left. \frac{\partial \varepsilon}{\partial {p}}\right\vert_{s} (\delta{p})^2 +  \frac{1}{2 (\varepsilon + {p})}\left.\frac{\partial \varepsilon}{\partial s}\right\vert_p \left.\frac{\partial {p}}{\partial s}\right\vert_{\alpha_q} (\delta s)^2 +  \left. \frac{\partial\alpha_q}{\partial {p}}\right\vert_s\delta {p}\,\delta\, n_q^{\mu} \lambda_\mu \nonumber\\
&+ \left. \frac{\partial \alpha_q}{\partial s}\right\vert_{p}\delta s\,\delta n_q^{\mu} \lambda_\mu +\delta\Pi \,\delta u^{\mu} \lambda_\mu +\delta \pi^{\mu\nu}\delta u_{\nu} \lambda_\mu + \frac{1}{2} \left( \beta_\Pi\delta \Pi\delta \Pi+ \beta_\pi\delta {\pi}^{\mu\nu}\delta {\pi}_{\mu\nu}+ \sum_q\beta_{n}^{qq'}\delta n_q^\mu\delta {n}_\mu^{q'}\right)\nonumber\\
&+ \sum_q \left(\delta_{n\Pi}^q \lambda_\mu\delta n^\mu\delta \Pi +\delta_{n\pi}^{q} \lambda_\mu\delta n_\nu^q\delta \pi^{\mu\nu}\right)\,,
\label{eq:Lyabunov-density}
\end{align}
here $\lambda^\mu = \frac{\Delta^{\mu\nu} n_\nu}{n_c u^c}$. In order to prove the positivity of the energy functional, we first write it in terms of linearly independent combinations. Then, we assure these combinations are positive definite by completing the squares of all coupled terms.

To do so, we first write the energy functional Eq.\ (\ref{eq:Lyabunov-density}) in the form of $k\times k$ dimensional matrix where k is the dimensionality of the phase space or state configurations. Thus, proving positivity of this matrix is a direct proof of stability of the dynamical system.

\begin{center}
$\left( \begin{array}{c}
     \delta u^{\mu}\\
     \delta p\\
     \delta s\\
     \delta n_q^{\mu}\\
     \delta \Pi_{\mu\nu}\\
     \delta \pi\\
\end{array}\right)^T\left( \begin{array}{ccc|ccccc}
      \frac{(\varepsilon+ p )}{2 }& \frac{\lambda_\mu}{2} &0&0& \frac{\lambda_\mu}{2}& \frac{\lambda_\nu}{2}\\
      \frac{\lambda_\mu}{2}& \frac{1}{(\varepsilon+ p )} \left. \frac{\partial \varepsilon}{\partial p}\right\vert_{s}&0& \frac{\lambda_\mu}{2} \left. \frac{\partial \alpha_q}{\partial p}\right\vert_{s}&0& 0\\
      0& 0 &\left. \frac{1}{(\varepsilon + p) } \left.\frac{\partial \varepsilon}{\partial s}\right\vert_{p} \frac{\partial p}{\partial s}\right\vert_{\alpha_q}&\frac{\lambda_\mu}{2} \left. \frac{\partial \alpha_q}{\partial s}\right\vert_{p}&0&0&\\
      \\
      \hline
      \\
      0&\frac{\lambda_\mu}{2} \left. \frac{\partial \alpha_q}{\partial p}\right\vert_{s}&\frac{\lambda_\mu}{2} \left. \frac{\partial \alpha_q}{\partial s}\right\vert_{p}&\beta_n^{qq'}&\frac{ \lambda_\mu}{2}\delta_{n\Pi}^q& \frac{\lambda_\mu}{2}\delta_{n\pi}^q\\
    \frac{ \lambda_\mu}{2}&0&0&\delta_{n\Pi}^q \frac{ \lambda_\mu}{2}&\beta_\Pi&0\\
     \frac{ \lambda_\nu}{2}&0&0 &\delta_{n\pi}^q \frac{\lambda_\mu}{2}&0&\beta_\pi\\
\end{array}\right)\left( \begin{array}{c}
     \delta u^{\mu}\\
     \delta p\\
     \delta s\\
     \delta n_q^{\mu}\\
     \delta \Pi_{\mu\nu}\\
     \delta \pi\\
\end{array}\right)\vspace{0.5cm}$\\
\end{center}

For a symmetric matrix, the proof that the determinants of all upper left matrices are positive suffice to prove its positivity. 
 
The conditions relate in a  non-trivial way and can be sorted into: expressions combining thermodynamics quantities such as $(\varepsilon, P)$, thermodynamics potentials and derivatives $(\alpha_q, c_s^2)$ as well as dissipative corrections $(\Pi,\pi_{\mu\nu})$, and transport coefficients $(\beta_\Pi, \beta_\pi, ...)$.  Here we only list the final conditions and discuss their physical implications. a detailed derivation is done in Sec. \ref{app:stability}. We classify the conditions as follow:
\begin{itemize}
\item Thermodynamics constraints:
\end{itemize}
\begin{align}\label{eqn:c1c2}
c1:\;\frac{1}{\varepsilon + p} \left.\frac{\partial \varepsilon}{\partial p}\right\vert_{s} \geq 0\,; \qquad\qquad\qquad c2:\;\frac{1}{\varepsilon + p} \left. \frac{\partial \varepsilon}{\partial s}\right\vert_{p} \left. \frac{\partial p}{\partial s}\right\vert_{\alpha_q}\geq 0\,.  
\end{align}
where Eq. (\ref{eqn:c1c2}) is the thermodynamic constraint on the speed of sound since $c_s^2=\left.\frac{\partial p}{\partial \varepsilon}\right\vert_{s} $. 
For most regions of the QCD phase diagram, it is reasonable to assume that the enthalpy $w=\varepsilon + p\geq 0$ and $c_s^2\geq 0$ are positive such that these relations in Eq. (\ref{eqn:c1c2}) are rather trivially satisfied.  However, if one passes through a metstable phase by spinodal decomposition then it is possible to reach a regime where $c_s^2<0$ or $w<0$.
The only configuration where stability is maintained while $c_s^2$ is allowed to be negative is if the enthalpy is negative i.e $\varepsilon + p < 0$, which implies that $p<0$ and $|\varepsilon| < |p|$ since the energy density must be positive. In this case, the speed of sound is imaginary i.e. $c_s \in \Im$ and $c_s^2<0$. The consequence of a negative pressure and $c_s^2$, is that it would imply that   $\left. \frac{\partial \varepsilon}{\partial s}\right\vert_{p} \left. \frac{\partial p}{\partial s}\right\vert_{\alpha_q}<0$ must be negative.  Conversely, if we know that $c_s^2>0$, then $w>0$ such that it implies $\left. \frac{\partial \varepsilon}{\partial s}\right\vert_{p} \left. \frac{\partial p}{\partial s}\right\vert_{\alpha_q}>0$ must be positive. Then, $\left. \frac{\partial \varepsilon}{\partial s}\right\vert_{p} \left. \frac{\partial p}{\partial s}\right\vert_{\alpha_q}$ is a thermodynamic condition on heat conduction.
\begin{itemize}
\item Transport constraints:
\end{itemize}
\begin{align}\label{eqn:TCcon}
c3:\;\beta_\Pi \geq 0\,, \qquad\qquad\qquad c4:\;\beta_\pi \geq 0\,; \qquad\qquad  
\end{align}
where $\lambda$ is the norm of $\lambda^\mu$. The first two conditions constrain the shear and bulk viscosities to be positive.
\begin{itemize}
\item Constraints connecting thermodynamics with transport: 
\end{itemize}
These condition have a more complicated form; however, one can use the previous inequalities to simplify them. For example,
\begin{align}
 c5:\;\frac{\varepsilon+p}{\lambda^2} \geq \frac{(\delta_{n\pi}^q+1)^2}{2\beta_\pi} -\frac{\rho_q^2 \beta_{n}^{qq'}}{\lambda^2}  
\end{align}
that nicely ties together a relation between the enthalpy $w=\varepsilon+p$, diffusion matrix $\beta_{n}^{qq'}$, and the coupling of diffusion to the shear stress $\delta_{n\pi}^q$. Since the charge density and the coupling coefficients are squared and $\beta_\pi$ is constrained to be positive by Eq.\ (\ref{eqn:TCcon}), their sign does not directly impact the matrix elements but their magnitude does. For example for $w>0$, the constraint c5 allows for a negative $\beta_{n}^{qq'}$ as long as the total $\frac{(\delta_{n\pi}^q+1)^2}{2\beta_\pi} -\frac{\rho_q^2 \beta_{n}^{qq'}}{\lambda^2}$ is less than the entropy $w$.

In contrast, if $w<0$, then we have:
\begin{align}\label{eqn:c5_wneg}
 \textrm{If } \varepsilon+p<0,\; \textrm{then}\quad \frac{(\delta_{n\pi}^q+1)^2}{2\beta_\pi} < \rho_q^2 \beta_{n}^{qq'}\, \textrm{and}\;   \beta_{n}^{qq'}>0
\end{align}
since the term $\frac{(\delta_{n\pi}^q+1)^2}{2\beta_\pi}$ is positive definite due to c4. Then, Eq.\ (\ref{eqn:c5_wneg}) implies that for a meta-stable region we must have $\beta_{n}^{qq'}>0$. 
Indeed, $\beta_{n}^{qq'}$ will have a dependence on the choice of the hyper-surface through the norm $\lambda^2$; however, in the above relation, we used the most stringent constraint by setting $\lambda^2 \rightarrow 1$.

The next condition states that the relation between transport and diffusion coefficients is constrained by the thermodynamics such that if we look at
\begin{align}\label{eqn:c6}
c6:\;\frac{\beta_{n}^{qq'}}{2\lambda^2} - \frac{({\delta_{n\pi}^q})^2}{\beta_\pi}-  \frac{(\delta_{n\Pi}^q)^2}{\beta_\Pi} \geq { \frac{\varepsilon + p}{4} \left. \frac{\partial p}{\partial \varepsilon}\right\vert_{s} \Bigg(\left. \frac{\partial {\alpha_q}}{\partial p}\right\vert_{s}\Bigg)^2 }- T^2 (\varepsilon+p) \left. \frac{\partial s}{\partial \varepsilon}\right\vert_{p} \left. \frac{\partial s}{\partial p}\right\vert_{\alpha_q} \Bigg(\left. \frac{\partial {\alpha_q}}{\partial s}\right\vert_{p}\Bigg)^2\,.
\end{align}
One finds can find additional constraints from c6 in very specific scenarios.  

{\bf Scenario 1} the speed of sound is $c_s^2>0$ and the right-hand side of Eq.\ (\ref{eqn:c6}) is positive:
\begin{align}\label{eqn:c6_cs2pos}
\textrm{if}\; c_s^2>0\quad \textrm{and}\quad & \frac{1}{4} \left. \frac{\partial p}{\partial \varepsilon}\right\vert_{s} \Bigg(\left. \frac{\partial {\alpha_q}}{\partial p}\right\vert_{s}\Bigg)^2  \geq \underbrace{T^2  \left. \frac{\partial s}{\partial \varepsilon}\right\vert_{p} \left. \frac{\partial s}{\partial p}\right\vert_{\alpha_q} \Bigg(\left. \frac{\partial {\alpha_q}}{\partial s}\right\vert_{p}\Bigg)^2}_{>0\textrm{ from c2}}\nonumber\\
\textrm{then}\quad& \frac{\beta_{n}^{qq'}}{2\lambda^2} \geq \frac{({\delta_{n\pi}^q})^2}{\beta_\pi}+ \frac{(\delta_{n\Pi}^q)^2}{\beta_\Pi}\,, 
\end{align}
that ensures the positivity of the right hand side of the inequality in Eq.\ (\ref{eqn:c6}). 

{\bf Scenario 2} the speed of sound is $c_s^2>0$ and the the right-hand side of Eq.\ (\ref{eqn:c6}) is negative:
\begin{align}\label{eqn:c6_cs2neg}
\textrm{if}\; c_s^2>0\quad \textrm{and}\quad & \frac{1}{4} \left. \frac{\partial p}{\partial \varepsilon}\right\vert_{s} \Bigg(\left. \frac{\partial {\alpha_q}}{\partial p}\right\vert_{s}\Bigg)^2  < T^2  \left. \frac{\partial s}{\partial \varepsilon}\right\vert_{p} \left. \frac{\partial s}{\partial p}\right\vert_{\alpha_q} \Bigg(\left. \frac{\partial {\alpha_q}}{\partial s}\right\vert_{p}\Bigg)^2\nonumber\\
\textrm{then}\quad& \frac{\beta_{n}^{qq'}}{2\lambda^2} \;\textrm{may be positive or negative}
\end{align}
that ensures the positivity of the right hand side of the inequality in Eq.\ (\ref{eqn:c6}). 

{\bf Scenario 3} for  $c_s^2<0$ we know from Eq.\ (\ref{eqn:c5_wneg}) that $\beta_{n}^{qq'}>0$ such that there is always a possibility for the left-hand side of Eq.\ (\ref{eqn:c6}) to be positive or negative, depends  on if $\beta_{n}^{qq'}$ is larger or smaller than $\frac{({\delta_{n\pi}^q})^2}{\beta_\pi}+ \frac{(\delta_{n\Pi}^q)^2}{\beta_\Pi}$.  If the right-hand side of Eq.\ (\ref{eqn:c6}) is negative, we do not have any additional constraints.  However, if the right-hand side of Eq.\ (\ref{eqn:c6}) is positive, then we find that
\begin{eqnarray}
\textrm{if}\; c_s^2<0;\quad\textrm{and} &\quad& \frac{1}{4} \left. \frac{\partial p}{\partial \varepsilon}\right\vert_{s} \Bigg(\left. \frac{\partial {\alpha_q}}{\partial p}\right\vert_{s}\Bigg)^2  \geq T^2  \left. \frac{\partial s}{\partial \varepsilon}\right\vert_{p} \left. \frac{\partial s}{\partial p}\right\vert_{\alpha_q} \Bigg(\left. \frac{\partial {\alpha_q}}{\partial s}\right\vert_{p}\Bigg)^2\\
\textrm{then}&\quad& \frac{\beta_{n}^{qq'}}{2\lambda^2} \geq \frac{({\delta_{n\pi}^q})^2}{\beta_\pi}+ \frac{(\delta_{n\Pi}^q)^2}{\beta_\Pi}\,, 
\end{eqnarray}
where we have then find that regardless of the sign of $c_s^2$, we can assume $\beta_{n}^{qq'}>0$ if Scenario 1 or Scenario 3 are fulfilled. However, it is possible for $\beta_{n}^{qq'}<0$ in Scenario 2.

Our last condition is the most complex and comes from the coupling between the charge current and the flow velocity which states the following inequality
\begin{align}
c7:\;&(\varepsilon + p) \bigg(\frac{1}{4} \left.\frac{\partial p}{\partial \varepsilon}\right\vert_{s}- \frac{1}{2\lambda^2}\bigg) 
-\frac{1}{\beta_\Pi} -{\frac{2}{3\beta_\pi} } \nonumber\\
&\geq \frac{\{ - \frac{\delta_{n\Pi}^q}{\beta_\Pi}-\frac{\delta_{n\pi}^q}{\beta_\pi}-\frac{\varepsilon + p}{2} \left.\frac{\partial p}{\partial \varepsilon}\right\vert_{s} \left.\frac{\partial {\alpha_q}}{\partial p}\right\vert_{s}\}^2}{\lambda^2\,\{\frac{\beta_{n}^{qq'}}{2\lambda^2} - \frac{({\delta_{n\pi}^q})^2}{\beta_\pi}-  \frac{(\delta_{n\Pi}^q)^2}{\beta_\Pi} -{ \frac{\varepsilon + p}{4} \left.\frac{\partial p}{\partial \varepsilon}\right\vert_{s} \Bigg(\left.\frac{\partial {\alpha_q}}{\partial p}\right\vert_{s}\Bigg)^2 }+ T^2 (\varepsilon+p) \left.\frac{\partial s}{\partial \varepsilon}\right\vert_{p} \left.\frac{\partial s}{\partial p}\right\vert_{\alpha_q} \Bigg(\left.\frac{\partial {\alpha_q}}{\partial s}\right\vert_{p}\Bigg)^2\}} 
\end{align}
Note that we have shown the constraints on the positivity of the denominator on the right hand side in the previous condition.
 
The constraints $c4, c5, c6, c7$ depend on the space-like vector $\lambda^2= \lambda_\mu \lambda^\mu$ where different choices of the hyper-surface will affect the particular value of this vector. As we show in the text, one can enforce the strongest constraint by setting $\lambda^2=1$. We finally note that the different structure of the conditions in the Landau frame compared to the Ekart frame \cite{Hiscock:1983zz, Olson:1990rzl} comes from the fact that charges could diffuse independently and are not restricted to co-move with the flow velocity. This creates a linear dependence between the fluctuations of $\delta u^\mu$ and $\delta n^\mu_q$, which we considered when deriving the constraints. While the general structure of the conditions may seem different, it has been recently shown in \cite{Gavassino:2022roi}  that one can map from one set of conditions to the other by a change of frame argument. A consequence of this effect for example is that it relaxes the constraint on the diffusion matrix c5 to allow for negative diffusion in some scenarios. One might still recover the positive definiteness of $\beta_{qq'}$ in the case of ideal BSQ hydrodynamics where the charge fluctuations co-move with the energy flow. 
\section{Discussion and Conclusions}
\label{sec:conclude}
In this paper, we derived the relativistic viscous hydrodynamics equations of motion with multiple conserved charges using the phenomenological approach from Israel-Stewart. For the first time, we were able to recover the DNMR $(K_n^{(1)} R_n^{(1)})$ terms from the phenomenological approach from Israel-Stewart approach by adding new terms to the entropy production that exactly cancel but that enter different dissipative equations of motion e.g. the  bulk pressure and the charge current equations.  This theoretical description is relevant for low energy heavy ion collisions and the search for critical point/first-order phase transition and may be of interest for neutron star mergers as well \cite{Most:2021zvc,Most:2022yhe}.

After our equations of motion were derived, we performed a stability analysis where the conditions are sufficient for hydrodynamic stability and are more general than a mode analysis. As the work in \cite{Gavassino:2021kjm} show, using the energy functional method, the conditions of thermodynamic stability at linear order directly imply the linear causality. This means that the conditions we derive here stand as penchmarks for stability as well as causality of the BSQ Israel-Stewart theory in the Landau frame. 

In this analysis, we find 7 stability constraints that either affect the thermodynamic variables, the transport coefficients, or a mixature of the two.We explore the possibility of the the meta-stable region across a first-order phase transition where $c_s^2<0$ is possible and find that we should still maintain hydrodynamic stability within that region as long as the enthalpy is also negative and the Navier Stokes transport coefficients for shear viscosity, bulk viscosity, and charge diffusion are positive definite. Interestingly enough, under very specific conditions outside of the meta-stable region (i.e. $c_s^2>0$) we can allow for a negative charge diffusion.

At large baryon densities the role of transport coefficients are potentially even more crucial than at $\mu_B=0$. Unfortunately, due to the fermion sign problem we have no guidance from first principle lattice QCD calculations on their behavior. That being said studies of critical scaling found that at the critical point the  bulk viscosity scales with the cubic power of the correlation \cite{Moore:2008ws,Monnai:2016kud,Martinez:2019bsn} and is already large because of the dip in $c_s^2\rightarrow 0$ \cite{Dore:2020jye,Dore:2021sbl}.  Across a first-order phase transition, one expects a jump in the transport coefficients \cite{Soloveva:2020hpr,Grefa:2022sav}  (although, for a meta stable region things may be more complicated and we are not aware of papers exploring that possibility yet in the context of heavy-ion collisions).  Thus, any constraints that are possible on their behavior are extremely important for simulations of heavy-ion collisions.

These conditions imply the causality of the solution at least in the linear regime. As  shown in Sec. \ref{subsec:conditions}, the stability criteria creates a direct correspondence between the EoS and the microscopic non-equilibrium physics. The distinction from the non-linear causality constraints derived in \cite{Bemfica:2020xym} is that our stability constraints are expressed in terms of thermodynamics and transport coefficients only while the conditions for causality include the magnitude of the dissipative fluxes. In the IS formalism, the transport coefficients set the magnitude of those contributions, so a constraint on the transport coefficient already constraints the relevant flux term.

One surprising outcome from this analysis is that we find that the terms that only appear in DNMR but not previous works on Israel-Stewart have zero contributions to the entropy current such that they do not affect thermodynamic stability. From the second law of thermodynamics, irreversible processes increase the entropy production. Since those terms do not increase the entropy, they might not transfer any energy and momentum on the hyper-surface; and therefore, they carry no information. The question then is what role they play, and how do they arise from the Boltzmann equation in the DNMR framework. 

For the reasons mentioned above, the extra DNMR terms would only be constrained from non-linear causality analysis. Previous work have found that these extra terms better handle far-from-equilibrium effects that can occur at the critical point \cite{Dore:2020jye} (i.e. the inclusion of these terms more quickly reached an attractor) but that work also never found that the lack of these terms lead to any thermodynamic instabilities.  
Thus, it appears that these extra terms shift out-of-equilibriumness between different dissipitative currents.  For instance, imagine a large peak in the bulk viscosity.  Then these terms could shift some the enhancement over to the shear stress tensor or charge currents to avoid issues with a large bulk viscosity such that the out-of-equilibriumness is more balanced between different disspitive currents.  We leave a more concrete study of these terms to an upcoming future work.
\section{Acknowledgements:}
We thank G.~S.~Denicol and J.~Noronha for helpful comments. J.N.H, T.D., D.A. are supported by the US-DOE Nuclear Science Grant No. DE-SC0020633.

\bibliography{inspire}
\newpage
\appendix
\section{shear-bulk coupling}
In this appendix, we calculate the divergence of the zero entropy current term
\begin{align}
    \partial_\mu (\delta_{\Pi \pi} u_\nu \Pi \pi^{\mu\nu}) &=  \delta_{\Pi \pi}\Pi \pi^{\mu\nu} (\partial_\mu u_\nu)+  u_\nu \Pi {\pi^{\mu\nu}} (\partial_\mu{\delta}_{\Pi\pi}) +\delta_{\Pi \pi} u_\nu \pi^{\mu\nu} (\partial_\mu{\Pi}) +\delta_{\Pi\pi} \Pi u_\nu (\partial_\mu{\pi}^{\mu\nu})\nonumber\\
    &= \delta_{\Pi \pi}\Pi \pi^{\mu\nu} \partial_\mu u_\nu + \delta_{\Pi\pi} \Pi u_\nu \partial_\mu{\pi}^{\mu\nu}\nonumber\\
    &= \delta_{\Pi \pi}\Pi \pi^{\mu\nu} \nabla_\mu u_\nu- \delta_{\Pi \pi}\Pi \pi^{\mu\nu} \nabla_\mu u_\nu 
\label{eqn:div-zero}
\end{align}
Where in going from second to third line in Eq. (\ref{eqn:div-zero}) we used the relation
\begin{eqnarray}
\partial_{\mu}\underbrace{\left(u_{\nu}\pi^{\mu\nu}\right)}_\textrm{=0}&=&u_{\nu}\partial_{\mu}\pi^{\mu\nu}+\pi^{\mu\nu}\partial_{\mu}u_{\nu}\nonumber
\label{eqn:parsub}
\end{eqnarray}
Which trivially gives
\begin{eqnarray}
u_{\nu}\partial_{\mu}\pi^{\mu\nu}&=&-\pi^{\mu\nu}\partial_{\mu}u_{\nu}\nonumber
\end{eqnarray}
And manipulated the relation
\begin{eqnarray}
\partial^{\mu}-\nabla^{\mu}&=&\partial_{\alpha}g^{\mu\alpha}-\Delta^{\mu\alpha}\partial_{\alpha}\nonumber\\
&=&\left[g^{\mu\alpha}-\Delta^{\mu\alpha}\right]\partial_{\alpha}\nonumber\\
&=&u^{\mu}u^{\alpha}\partial_{\alpha}\nonumber\\
&=&u^{\mu}D
\end{eqnarray}
such that
\begin{eqnarray}
\pi^{\mu\nu} \partial_\mu u_\nu &=& \pi^{\mu\nu}(u_\mu D + \nabla_\mu) u_\nu\nonumber\\
&=& \pi^{\mu\nu} \nabla^\mu u^\nu.
\end{eqnarray}

Back to Eq. (\ref{eqn:div-zero}), we see that therm $\nabla_{\mu}u_{\nu}$ can be divided into its symmetric $(\dots)$ and anti-symmetric $\{\dots\}$ part
\begin{eqnarray}
\nabla_{\mu}u_{\nu}&=& \frac{1}{2}\left(\nabla_{\mu}u_{\nu}+\nabla_{\nu}u_{\mu}\right)+\frac{1}{2}\left(\nabla_{\mu}u_{\nu}- \nabla_{\nu}u_{\mu}\right)\nonumber\\
&=&\sigma_{\mu\nu}+\frac{1}{3}\Delta_{\mu\nu}\theta+\Omega_{\mu\nu}
\end{eqnarray}
where
\begin{eqnarray}
\sigma_{\mu\nu}&\equiv&\frac{1}{2}\left(\nabla_{\mu}u_{\nu}+\nabla_{\nu}u_{\mu}-\frac{2}{3}\Delta_{\mu\nu}\theta\right)\nonumber\\
&=&\nabla_{(\mu}u_{\nu)}-\frac{1}{3}\Delta_{\mu\nu}\theta\label{eqn:sigmdef}\\
\theta&\equiv&\nabla_{\mu}u^{\mu}\\
\Omega_{\mu\nu}&\equiv&\frac{1}{2}\left(\nabla_{\mu}u_{\nu}-\nabla_{\nu}u_{\mu}\right)
\end{eqnarray}
where the shear stress tensor $\sigma_{\mu\nu}$ is symmetric, the expansion rate $\Theta$ is symmetric, and the vorticity $\Omega_{\mu\nu}$ is anti-symmetric. Inserting this symmetric form back into the term $\delta_{\Pi \pi}\Pi \pi^{\mu\nu} \nabla_\mu u_\nu$ will get us
\begin{align}
\delta_{\Pi \pi}\Pi \pi^{\mu\nu} \nabla_\mu u_\nu&= \delta_{\Pi \pi}\Pi \pi^{\mu\nu} \big( \sigma_{\mu\nu}+\frac{1}{3}\Delta_{\mu\nu}\theta+\Omega_{\mu\nu}\big)\nonumber\\
&= \delta_{\Pi \pi}\Pi \pi^{\mu\nu}  \sigma_{\mu\nu}
\label{eqn:zero-term}
\end{align}
where $\pi^{\mu\nu}\Omega_{\mu\nu}=0$ and $\frac{1}{3} \pi^{\mu\nu} \Delta_{\mu\nu}\theta=0$ by symmetry. This term is the new addition to the bulk and shear dynamical relaxation equations which arises from the relevant zero entropy current contribution.
\section{Lyapunov functional}
\label{app:stability}
To construct an energy functional which is quadratic in the perturbation fields, we start first by expressing the entropy current Eq. ($\ref{Eq:entropy-current}$) as ${S}^\mu = {S}^\mu_0 + {S}^\mu_1 + {S}^\mu_2$.

Then, one can express the fluctuations around the equilibrium state as
\begin{equation}
\delta {S}^\mu =\delta {S}^\mu_0 +\delta {S}^\mu_1 +\delta {S}^\mu_2.
\end{equation}
The perturbed fields are as follows
\begin{align}
\Tilde{\varepsilon} &= \varepsilon+\delta \varepsilon; \qquad\qquad\qquad\qquad\qquad\qquad \Tilde{T} = T +\delta T; \qquad\qquad\qquad\qquad\qquad\qquad \Tilde{\Pi}= \Pi+\delta \Pi\nonumber\\
\Tilde{P} &= p +\delta p;\qquad\qquad\qquad\qquad\qquad\qquad \Tilde{s} = s+\delta s;\qquad\qquad\qquad\qquad\qquad\qquad\Tilde{\pi}^{\mu\nu}= \pi^{\mu\nu}+\delta \pi^{\mu\nu} \nonumber\\
\Tilde{\rho_q} &= \rho_q +\delta \rho_q;\qquad\qquad\qquad\qquad\qquad\Tilde{\mu_q}= \mu_q+\delta \mu_q ;\qquad\qquad\qquad\qquad\qquad\qquad  \Tilde{n}^{\mu}_q= {n}^{\mu}_q +\delta {n}^{\mu}_q \nonumber\\
&\qquad\qquad\qquad\qquad\qquad\qquad\qquad\qquad\qquad\qquad\qquad\qquad\qquad\qquad\qquad\qquad \qquad\Tilde{u}^{\mu}= {u}^{\mu} +\delta {u}^{\mu}
\end{align}
for perturbations at $0^{th}$ order of the equilibrium entropy current
\begin{equation}
S^{\mu}_0= s u^{\mu},
\end{equation}
with 
\begin{equation}
s T =   \varepsilon + p - \sum_q \mu_q \,\rho_q, \qquad \qquad q\in\{B,S,Q\}\,.
\label{Eq:Gibbs}
\end{equation}
Therefore, the perturbed expression for this entropy becomes
\begin{align}
 S^{\mu}_0 +\delta S^{\mu}_0 &=  (s +\delta s) (u^{\mu} +\delta  u^{\mu})\nonumber\\
 S^{\mu}_0 +\delta S^{\mu}_0 &= s u^{\mu} + s\delta u^{\mu} +\delta s u^{\mu} +\delta s\delta u^{\mu}\nonumber\\
\delta S^{\mu}_0 &= s\delta u^{\mu} +\delta s u^{\mu} +\delta s\delta u^{\mu}.
\label{Eq:0thperturb}
\end{align}
while the perturbed form of the thermodynamic relation \ref{Eq:Gibbs} gives
\begin{align}
\delta s\, T &=  (\varepsilon + p - \sum_q \mu_q \,\rho_q - s T)+\delta \varepsilon +\delta p  - \sum_q (\mu_q \,\delta \rho_q +\delta \mu_q \,\rho_q +\delta \mu_q \,\delta \rho_q)  - s\delta T -\delta s\delta T\,,\nonumber\\
 &= \delta \varepsilon  +\delta p - \mu_q \,\delta \rho_q -\delta \mu_q \,\rho_q -  \delta \mu_q \,\delta \rho_q - s\delta T 
 - \delta s\delta T\,. 
\end{align}
The first term in parentheses go to zero by the identity Eq. (\ref{Eq:Gibbs}). Inserting the expression into Eq. (\ref{Eq:0thperturb}) and keeping only up to second order terms gives
\begin{align}
T\delta S^{\mu}_0 &=  \left[\varepsilon + p - \mu_q \,\rho\right]\delta u^{\mu} +  \left[\delta \varepsilon - \mu_q \,\delta \rho+\delta p -\delta \mu_q \,\rho - s\delta T- \delta \mu_q \,\delta \rho + \delta s\delta T\right] u^{\mu} \nonumber\\
& + \left[\delta \varepsilon - \mu_q \,\delta \rho+\delta p -\delta \mu_q \,\rho -  s\delta T \right]\delta u^{\mu}\,.
\end{align}
Finally eliminating the third term in the third square brackets using  the thermodynamics identity (Gibbs-Duhem) gives
\begin{align}
T\delta S^{\mu}_0 &= \left[\varepsilon + p - \mu_q \,\rho\right]\delta u^{\mu} + \left[\delta \varepsilon - \mu_q \,\delta \rho-\delta \mu_q \,\delta \rho + \delta s\delta T\right] u^{\mu} + \left[\delta \varepsilon - \mu_q \,\delta \rho\right]\delta u^{\mu}\,.
\label{eq:0th-pert}
\end{align}
Now, for the 1st order perturbations
\begin{align}
{S}^\mu_1 &= \alpha_q n^\mu_q\,, \qquad\qquad q\in{B,S,Q}\nonumber\\
\delta {S}^\mu_1 &= \alpha_q  n^\mu_q+ \alpha_q \delta n^\mu_q+ \delta \alpha_q \delta n^\mu_q+ \delta \alpha_q \delta n^\mu_q\,, 
\label{eq:1st-pert}
\end{align}
Finally, for the 2nd order perturbations
\begin{equation}
{S}^\mu_2 = - \frac{1}{2}u^\mu\left( \beta_\Pi{\Pi}^2 + \beta_\Pi {\pi}^{\mu\nu}{\pi}_{\mu\nu} + \sum_q\beta_{n}^{qq'} n_q^\mu{n}_\mu^{q'} \right) - \sum_q\left(\delta_{n\pi}^q n^\mu_q {\Pi} +\delta_{n\pi}^{q} n_\nu^q{\pi}^{\mu\nu} \right)\,,
\end{equation}
which then can be expanded as 
\begin{align}
\delta {S}^\mu_2 &= - \frac{1}{2} \left(u^\mu+\delta u^\mu\right) \left[\beta_\Pi\left(\Pi +\delta \Pi \right)^2 + \beta_\Pi \left({\pi}^{\mu\nu} +\delta {\pi}^{\mu\nu})({\pi}_{\mu\nu} +\delta {\pi}_{\mu\nu}\right)+ \sum_q\beta_{n}^{qq'} (n_q^\mu +\delta n_q^\mu)
( {n}_\mu^{q'}+\delta {n}_\mu^{q'})\right]\nonumber\\
&- \sum_q\left[\delta_{n\Pi}^q (n^\mu_q+\delta n_q^\mu)( \Pi+\delta \Pi) +\delta_{n\pi}^{q} (n_\nu^q+\delta n_\nu^q)({\pi}^{\mu\nu}+\delta {\pi}^{\mu\nu}) \right]\,.
\end{align}
At this point, we set all terms proportional to the currents $\{\Pi, \pi^{\mu\nu}, n^\mu_q; q\in \{B,S,Q\}\}$ to zero by definition of the equilibrium state and keep only terms at second order in the perturbations to arrive at
\begin{align}
\delta {S}^\mu_2 &= - \frac{1}{2}u^\mu \left( \beta_\Pi\delta \Pi\delta \Pi+ \beta_\Pi\delta {\pi}^{\mu\nu}\delta {\pi}_{\mu\nu}+ \sum_q\beta_{n}^{qq'}\delta n_q^\mu\delta {n}_\mu^{q'}\right)- \sum_q \left(\delta_{n\Pi}^q\delta n^\mu_q\delta \Pi +\delta_{n\pi}^{q}\delta n_\nu^q\delta \pi^{\mu\nu}\right).
\label{eq:2nd-pert}
\end{align}
To get the full perturbed non-equilibrium entropy, we add up all terms in Eqs. (\ref{eq:0th-pert}, \ref{eq:1st-pert}, and \ref{eq:2nd-pert}) to get 
\begin{align}
T\delta S^\mu_{non-eq}&= \left[\varepsilon + p - \sum_q \mu_q \,\rho \right]\delta u^{\mu} + \left[\delta \varepsilon - \sum_q (\mu_q \,\delta \rho-\delta \mu_q \,\delta \rho) + \delta s\delta T\right] u^{\mu}
+ \left[\delta \varepsilon- \sum_q \mu_q \,\delta \rho\right]\delta u^{\mu}\nonumber\\
&-\alpha_q \delta n_q^{\mu} - \frac{1}{2}u^\mu \left( \beta_\Pi\delta \Pi\delta \Pi+ \beta_\Pi\delta {\pi}^{\mu\nu}\delta {\pi}_{\mu\nu}+ \sum_q\beta_{n}^{qq'}\delta n_q^\mu\delta {n}_\mu^{q'}\right)
- \sum_q \left(\delta_{n\Pi}^q\delta n^\mu\delta \Pi +\delta_{n\pi}^{q}\delta n_\nu^q\delta \pi^{\mu\nu}\right)
\end{align}
The idea is that the energy-momentum and number should be conserved for all physical state configurations. These "zero flux contributions" are calculated using the next two expressions
\subsubsection{zero energy-momentum flux contribution}
We start from the perturbed form of the Landau condition, $u_{\mu} T^{\mu\nu}= \varepsilon u^{\nu}$, imposing that no fluctuations can propagate 
\begin{align}
(u_{\nu} +\delta u_{\nu} )(T^{\mu\nu}+\delta T^{\mu\nu})  &= (\varepsilon +\delta \varepsilon) (u^{\mu} +\delta u^{\mu})\,,\nonumber\\
u_{\nu} T^{\mu\nu}+u_{\nu}\delta T^{\mu\nu}+\delta u_{\nu} T^{\mu\nu} +\delta u_{\nu}\delta T^{\mu\nu} &= \varepsilon u^{\mu} + \varepsilon\delta u^{\mu} +\delta \varepsilon u^{\mu} +\delta \varepsilon \delta u^{\mu}\,, \nonumber\\
u_{\nu}\delta T^{\mu\nu}&=  \varepsilon\delta u^{\mu} +\delta \varepsilon u^{\mu} +\delta \varepsilon\delta u^{\mu}-\delta u_{\nu} T^{\mu\nu} -\delta u_{\nu}\delta T^{\mu\nu}\,,\nonumber\\
&= (\rm zfc)_{energy-momentum}\,.
\end{align}
Now we turn to the charge current(s) zero flux contribution. We start again from the definition of Landau frame of the currents expanded in perturbed form $N^{\mu}_q= \rho  u^{\mu} + n^{\mu}_q$. Note that this will impose conditions on each of the charges separately 
\begin{align}
N^{\mu}_q+\delta N^{\mu}_q &= (\rho_q +\delta \rho_q)(u^{\mu} +\delta u^{\mu}) +(n^{\mu}_q+\delta n^{\mu}_q)\,, \nonumber\\ 
&= \rho_q u_{\mu} + \rho_q\delta u_{\mu} +\delta \rho u_{\mu} +\delta \rho\delta u_{\mu} +n^{\mu}_q +\delta n^{\mu}_q\,,  \nonumber\\
\delta N^{\mu}_q&=  \rho_q \,\delta u_{\mu} +\delta \rho_q\, u_{\mu} +\delta \rho_q \,\delta u_{\mu} +\delta n^{\mu}_q \,.\nonumber\\
&= (\rm zfc)_{charges}
\end{align}
From energy flux contributions equation, we get
\begin{align}
{(\rm zfc)}_{T^{\mu\nu}} &=  \varepsilon\delta u^{\mu} +\delta \varepsilon u^{\mu} +\delta \varepsilon\delta u^{\mu}-\delta u_{\nu} T^{\mu\nu} -\delta u_{\nu}\delta T^{\mu\nu}\,, 
\end{align}
Similarly, from the number flux contributions equation one finds
\begin{align}
{(\rm zfc)}_{N_q^\mu} &=  \rho_q \,\delta u_{\mu} +\delta \rho_q\, u_{\mu} +\delta \rho_q \,\delta u_{\mu} +\delta n^{\mu}_q\,, \nonumber\\
\end{align}
and we obtain
\begin{align}
\mu_q\, \,\rho_q \delta u_{\mu} &= ({\rm zfc})  - \mu_q\,  \,\delta \rho_q u_{\mu} -\mu_q\, \, \delta \rho_q\delta u_{\mu} - \mu_q\,\delta n^{\mu}_q\,. \nonumber
\end{align}
The final expression is now
\begin{align}
T\delta S^\mu_{non-eq} &=({\rm zfc}) -\delta \varepsilon u^{\mu} -\delta \varepsilon\delta u^{\mu}+\delta u_{\nu} T^{\mu\nu} +\delta u_{\nu}\delta T^{\mu\nu}+p\delta u^{\mu} -  ({\rm zfc})  + {\mu_q\,\delta \rho} \, u_{\mu}\nonumber\\
&+{\mu_q\,\delta \rho} \,\delta u_{\mu} + {\mu_q\,\delta n^{\mu}_q} + \left[\delta \varepsilon - \mu_q \,\delta \rho_q- \delta \mu_q \,\delta \rho + \delta s\delta T\right] u^{\mu}+ \left[\delta \varepsilon- \mu_q \,\delta \rho\right]\delta u^{\mu}\nonumber\\
&-\mu_q\delta n_q^{\mu} -\delta \mu_q\delta n_q^{\mu}+ \frac{\mu_q\delta n_q^{\mu}\delta T}{T}- \frac{1}{2}u^\mu \left( \beta_\Pi\delta \Pi\delta \Pi+ \beta_\Pi\delta {\pi}^{\mu\nu}\delta {\pi}_{\mu\nu}+ \sum_q\beta_{n}^{qq'}\delta n_q^\mu\delta {n}_\mu^{q'}\right)\nonumber\\
&- \sum_q \left(\gamma_{n\Pi}^q\delta n^\mu\delta \Pi +\gamma_{n\pi}^{q}\delta n_\nu^q\delta \pi^{\mu\nu}\right)
\end{align}
To further simplify the above expression, we note first that the energy momentum tensor (in the mostly negative convention) is decomposed as
\begin{align}
T^{\mu\nu} = \varepsilon u^{\mu} u^{\nu} - (p + \pi)\delta^{\mu\nu} + \pi^{\mu\nu}    
\end{align}
Then by projecting into $\delta u^\mu$ and manipulating the relations $u^\nu\delta u_{\nu} =\delta u^{\nu}\delta u_{\nu}/2$, one finds
\begin{align}
\delta u_{\nu} T^{\mu\nu} &= \varepsilon\delta u_{\nu} u^{\mu} u^{\nu} - (p + \Pi)\delta u_{\nu}\delta^{\mu\nu} +\delta u _{\mu} \pi^{\mu\nu}\nonumber\\
\delta u_{\nu} T^{\mu\nu} &= \varepsilon\delta u_{\nu} u^{\mu} u^{\nu} - p\delta u_{\nu}\delta^{\mu\nu}  - \Pi\delta u_{\nu}\delta^{\mu\nu} +\delta u _{\nu} \pi^{\mu\nu}\nonumber\\
&= \varepsilon\delta u_{\nu} u^{\mu} u^{\nu} - p \delta u_{\nu} g^{\mu\nu}+ u_{\mu}  u_{\nu}p \delta u_{\nu} - \Pi\delta u^{\mu} +\delta u_{\nu} \pi^{\mu\nu}\nonumber\\
&= - (\varepsilon+ p ) \frac{\delta u_{\nu}\delta u^{\nu}}{2} u^{\mu}- p \delta u^{\mu}- \Pi\delta u^{\mu} +\delta u_{\nu} \pi^{\mu\nu}
\end{align}
Likewise
\begin{align}
T^{\mu\nu}+\delta T^{\mu\nu} &= (\varepsilon+\delta \varepsilon)u^{\mu} u^{\nu} - (p +\delta p)\delta^{\mu\nu} - (\Pi +\delta \pi)\delta^{\mu\nu} + (\pi^{\mu\nu}+\delta \pi^{\mu\nu} )\nonumber\\
&=\varepsilon u^{\mu} u^{\nu} +\delta \varepsilon u^{\mu} u^{\nu}- p\delta^{\mu\nu} -\delta p\delta^{\mu\nu} - \Pi\delta^{\mu\nu}-\delta \Pi\delta^{\mu\nu} + \pi^{\mu\nu}+\delta \pi^{\mu\nu} \nonumber\\
\end{align}
Expanding out all terms
\begin{align}
T^{\mu\nu}+\delta T^{\mu\nu} &= \varepsilon (u^{\mu} +\delta u^{\mu})( u^{\nu} +\delta u^{\nu}) +\delta \varepsilon(u^{\mu} +\delta u^{\mu})( u^{\nu} +\delta u^{\nu})- p (g^{\mu\nu}- (u^{\mu} +\delta u^{\mu})( u^{\nu} +\delta u^{\nu}))\nonumber\\
&- \Pi (g^{\mu\nu}- (u^{\mu} +\delta u^{\mu})( u^{\nu} +\delta u^{\nu})) + \pi^{\mu\nu}+\delta \pi^{\mu\nu}\nonumber\\
&= \varepsilon (u^{\mu}  u^{\nu} + u^{\mu}\delta u^{\nu} +\delta u^{\mu} u^{\nu} +\delta u^{\mu}\delta u^{\nu})+\delta \varepsilon (u^{\mu}  u^{\nu} + u^{\mu}\delta u^{\nu} +\delta u^{\mu} u^{\nu} +\delta u^{\mu}\delta u^{\nu}) \nonumber\\
&-  p g^{\mu\nu} - p (u^{\mu}  u^{\nu} + u^{\mu}\delta u^{\nu} +\delta u^{\mu} u^{\nu} +\delta u^{\mu}\delta u^{\nu})  -\delta p g^{\mu\nu} -\delta p (u^{\mu}  u^{\nu} +u^{\mu}\delta u^{\nu} \nonumber\\
& +\delta u^{\mu} u^{\nu} +\delta u^{\mu}\delta u^{\nu}) -  \Pi g^{\mu\nu} - \Pi (u^{\mu}  u^{\nu} + u^{\mu}\delta u^{\nu} +\delta u^{\mu} u^{\nu} +\delta u^{\mu}\delta u^{\nu})  -\delta \Pi g^{\mu\nu} \nonumber\\
&-\delta \Pi (u^{\mu}  u^{\nu} +u^{\mu}\delta u^{\nu}+\delta u^{\mu} u^{\nu} +\delta u^{\mu}\delta u^{\nu})+ \pi^{\mu\nu}+\delta \pi^{\mu\nu}. \end{align}
Canceling the unperturbed terms on both sides and discarding terms at third order in fluctuations gives
\begin{align}
\delta T^{\mu\nu} &= \varepsilon ( u^{\mu}\delta u^{\nu} +\delta u^{\mu} u^{\nu} +\delta u^{\mu}\delta u^{\nu})+\delta \varepsilon (u^{\mu}  u^{\nu} + u^{\mu}\delta u^{\nu} +\delta u^{\mu} u^{\nu}) \nonumber\\
& - p (u^{\mu}\delta u^{\nu} +\delta u^{\mu} u^{\nu} +\delta u^{\mu}\delta u^{\nu})  -\delta p g^{\mu\nu} -\delta p (u^{\mu}  u^{\nu} +u^{\mu}\delta u^{\nu} \nonumber\\
& +\delta u^{\mu} u^{\nu})  - \Pi ( u^{\mu}\delta u^{\nu} +\delta u^{\mu} u^{\nu} +\delta u^{\mu}\delta u^{\nu})  -\delta \Pi g^{\mu\nu} \nonumber\\
&-\delta \Pi (u^{\mu}  u^{\nu} +u^{\mu}\delta u^{\nu}+\delta u^{\mu} u^{\nu} )+\delta \pi^{\mu\nu}
\end{align}
Finally, we project with $\delta u_{\nu}$ keeping in mind that it preserves all symmetry arguments as the flow field $u_{\nu}$ and cancelling all third order terms
\begin{align}
\delta u_{\nu}\delta T^{\mu\nu} &= \varepsilon ( u^{\mu}\delta u^{\nu}\delta u_{\nu} +\delta u^{\mu}\delta u_{\nu}  u^{\nu})+\delta \varepsilon (u^{\mu}  u^{\nu}\delta u_{\nu}  )+ p (u^{\mu}\delta u^{\nu}\delta u_{\nu} +\delta u^{\mu} u^{\nu}\delta u_{\nu}  )  \nonumber\\
&-\delta p g^{\mu\nu}\delta u_{\nu}  +\delta p (u^{\mu}  u^{\nu}\delta u_{\nu} )- \Pi ( u^{\mu}\delta u^{\nu}\delta u_{\nu} +\delta u^{\mu} u^{\nu}\delta u_{\nu})  -\delta \Pi\delta u^{\mu} \nonumber\\
&-\delta \Pi (u^{\mu}  u^{\nu}\delta u_{\nu} +u^{\mu}\delta u^{\nu}\delta u_{\nu}+\delta u^{\mu} u^{\nu}\delta u_{\nu} )+\delta u_{\nu}\,\delta \pi^{\mu\nu} \nonumber\\
&= \varepsilon u^{\mu}\delta u^{\nu}\delta u_{\nu}+ p u^{\mu}\delta u^{\nu}\delta u_{\nu} -\delta p\delta u^{\mu} - \Pi u^{\mu}\delta u^{\nu}\delta u_{\nu} -\delta \Pi\delta u^{\mu} +\delta u_{\nu}\,\delta \pi^{\mu\nu} 
\end{align}
to arrive at 
\begin{align}
\delta u_{\nu} T^{\mu\nu} +\delta u_{\nu}\delta T^{\mu\nu}&=   - (\varepsilon+ p ) \frac{\delta u_{\nu}\delta u^{\nu}}{2} u^{\mu}- p \delta u^{\mu}+ \varepsilon u^{\mu}\delta u^{\nu}\delta u_{\nu}+ p u^{\mu}\delta u^{\nu}\delta u_{\nu} -\delta p\delta u^{\mu} \nonumber\\
&- \Pi\delta u^{\mu} +\delta u_{\nu} \pi^{\mu\nu}- \Pi u^{\mu}\delta u^{\nu}\delta u_{\nu} -\delta \Pi\delta u^{\mu} +\delta u_{\nu}\,\delta \pi^{\mu\nu}\nonumber\\ 
&=  (\varepsilon+ p - \Pi ) \frac{\delta u_{\nu}\delta u^{\nu}}{2} u^{\mu}  - (p +\delta p + \Pi)\delta u^{\mu} + (\pi^{\mu\nu} +\delta \pi^{\mu\nu})\delta u_{\nu}
\end{align}
The final step in the derivation is to subtract those terms which will provide us with the Lyabunov functional 
\begin{align}
E^{\mu} &= ({\rm zfc}) -\delta S^\mu_{non-eq}
\end{align}
Before we go further we note that some of our fluctuations are not independent from the others. So first we rewrite all quantities in linear combinations of independent ones to make sure our basis is complete and independent. 
First we need to simplify 
\begin{align}
\delta \mu_q \,\delta \rho_q + \delta s\delta T    
\end{align}
\begin{align}
ds= \frac{1}{T} d\varepsilon - \frac{\mu_q}{T} d\rho_q    
\end{align}
giving
\begin{align}
ds= \frac{1}{T} d\varepsilon - \frac{\mu_q}{T} d\rho_q    
\end{align}
\begin{align}
\delta \mu_q \,\delta \rho_q + \delta s\delta T = \frac{1}{\varepsilon + p} \left[\left. \frac{\partial \varepsilon}{\partial p}\right\vert_{s} (\delta p)^2 + \left. \frac{\partial \varepsilon}{\partial s}\right\vert_{p} \left.\frac{\partial p}{\partial s}\right\vert_{\mu_q/T} (\delta s)^2 \right] 
\end{align}
\begin{align}
\delta \frac{\mu_q}{T} = \left. \frac{\partial}{\partial p}\frac{\mu_q}{T}\right\vert_s\delta p + \left.\frac{\partial}{\partial s}\frac{\mu_q}{T}\right\vert_p\delta s
\end{align}
%
\section{Stability constraints}
The energy functional in Eq. (\ref{eq:Lyabunov}) is written as function of fluctuations including couplings which introduce nonlinear dependence between different fluctuations. In order to prove the positivity of the energy functional, we first write it in linearly independent combinations. Then, we assure these combinations are positive definite by completing the squares of all coupled terms. 

I will start first by hydrodynamics fluctuations terms which are coupled to thermodynamics
\begin{align}
e&=- (\varepsilon+ p) \frac{\delta u_{\nu} \delta u^{\nu}}{2 } + {\delta p \, \delta u^{\mu}} \lambda_\mu \pm \delta\Pi\, \delta u^{\mu} \lambda_\mu+ \delta \pi^{\mu\nu} \delta u_{\nu} \lambda_\mu+ {\frac{1}{\varepsilon + p}  \left. \frac{\partial \varepsilon}{\partial p}\right\vert_{s} (\delta p)^2} \nonumber\\
&+  {\frac{1}{\varepsilon + p} \left. \frac{\partial \varepsilon}{\partial s}\right\vert_{p} \left.\frac{\partial p}{\partial s}\right\vert_{\mu_q/T} (\delta s)^2} + {\sum_q \left. \frac{\partial\alpha_q}{\partial p}\right\vert_s \lambda_\mu \delta n_q^{\mu} \delta p} + {\sum_q \left. \frac{\partial \alpha_q}{\partial s}\right\vert_p \lambda_\mu \delta n_q^{\mu} \delta s} + \frac{\beta_\Pi}{2} \delta \Pi \delta \Pi\nonumber\\
&+ \frac{\beta_\pi}{2} \delta {\pi}^{\mu\nu}\delta {\pi}_{\mu\nu}+ \sum_q \frac{\beta_{n}^{qq'}}{2}  \delta n_q^\mu \delta {n}_\mu^{q'}+ \sum_q \delta_{n\pi}^q \lambda_\mu \delta n^\mu \delta \Pi + \sum_q \delta_{n\pi}^{q} \lambda_\mu \delta n_\nu^q \delta \pi^{\mu\nu}
\label{eq:energy}
\end{align}
Since we cannot constrain the magnitude of perturbations or how they propagate, we need to rewrite them in a quadratic form to be able to extract the conditions. Starting from the coupling between $(\delta u^\mu, \delta n_q^\mu, \delta p)$ perturbations, one gets
\begin{align}
e_{u^\mu,n_q^\mu,p}&=\delta p\,\delta u^{\mu} \lambda_\mu + \frac{1}{\varepsilon + p}  \left.\frac{\partial \varepsilon}{\partial p}\right\vert_{s} (\delta p)^2 + \sum_q \left. \frac{\partial\alpha_q}{\partial p}\right\vert|_s \lambda_\mu\delta n_q^{\mu}\delta p\nonumber\\
&=\frac{1}{\varepsilon + p} \left. \frac{\partial \varepsilon}{\partial p}\right\vert_{s} \left[\delta p + \frac{\varepsilon + p}{2} \left. \frac{\partial p}{\partial \varepsilon}\right\vert_{s}\left(\delta u^\mu \lambda_\mu + \left. \frac{\partial\alpha_q}{\partial p}\right\vert_s\delta n_q^\mu \lambda_\mu\right)\right]^2 - \frac{\varepsilon + p}{4} \left.\frac{\partial p}{\partial \varepsilon}\right\vert_{s}(\delta u^\mu \lambda_\mu)^2\nonumber\\
&  - \frac{\varepsilon + p}{4} \left. \frac{\partial p}{\partial \varepsilon}\right\vert_{s} \Bigg(\left. \frac{\partial {\alpha_q}}{\partial p}\right\vert_{s}\Bigg)^2 (\delta n_q^\mu \lambda_\mu) (\delta n_{q'}^\mu \lambda_\mu) - \frac{\varepsilon + p}{2} \left. \frac{\partial p}{\partial \varepsilon}\right\vert_{s} \left.\frac{\partial {\alpha_q}}{\partial p}\right\vert_{s} (\delta u^\mu \lambda_\mu)(\delta n_q^\mu \lambda_\mu).
\label{eq:unp}
\end{align}
We follow the same procedure for the couplings between $({\delta u^\mu, \delta n_q^\mu, \delta s})$ perturbations to get
\begin{align}
e_{n_q^\mu,s}&= T\left. \frac{\partial {\alpha_q}}{\partial s}\right\vert_{p}\delta s\,\delta n_q^{\mu} \lambda_\mu + \frac{1}{\varepsilon + p}  \left. \frac{\partial \varepsilon}{\partial s}\right\vert_{p} \left. \frac{\partial p}{\partial s}\right\vert_{\alpha_q} (\delta s)^2 \nonumber\\
&=\frac{1}{\varepsilon + p} \left. \frac{\partial \varepsilon}{\partial s}\right\vert_{p} \left. \frac{\partial p}{\partial s}\right\vert_{\alpha_q} \left[\delta s +  T (\varepsilon+p) \left. \frac{\partial s}{\partial \varepsilon}\right\vert_{p} \left. \frac{\partial s}{\partial p}\right\vert_{\alpha_q} \left.\frac{\partial {\alpha_q}}{\partial s}\right\vert_{p} (\delta n_q^\mu \lambda_\mu) \right]^2-  T^2 (\varepsilon+p) \nonumber\\
&\left. \frac{\partial s}{\partial \varepsilon}\right\vert_{p} \left.\frac{\partial s}{\partial p}\right\vert_{\alpha_q} \Bigg(\left. \frac{\partial {\alpha_q}}{\partial s}\right\vert_{p}\Bigg)^2 (\delta n_q^\mu \lambda_\mu)(\delta n_{q'}^\mu \lambda_\mu)
\end{align}
At this point we get a few linear independent combinations of the thermodynamics quantities with the flow and charge currents fields. We continue with coupled fluctuations among the pure hydrodynamics fields starting with the bulk fluctuations $({\delta \Pi,\delta u^{\mu},\delta n^{\mu}})$
\begin{align}
e_{\Pi,u^{\mu},n^{\mu}}&=  \frac{\beta_\Pi}{2} \delta \Pi \delta \Pi +\delta\Pi\, \delta u^{\mu} \lambda_\mu  + \sum_q \delta_{n\Pi}^q \lambda_\mu \delta n^\mu \delta \Pi \nonumber\\
&= \frac{\beta_\Pi}{2} \left[ \delta \Pi + \frac{1}{\beta_\Pi} \delta u^{\mu} \lambda_\mu \pm \frac{\delta_{n\Pi}^q}{\beta_\Pi} \delta n^{\mu}_q \lambda_\mu\right ]^2 - \frac{1}{\beta_\Pi} (\delta u^{\mu} \lambda_\mu)^2 - \frac{(\delta_{n\Pi}^q)^2}{\beta_\Pi} (\delta n^{\mu}_q \lambda_\mu)^2
- \frac{\delta_{n\Pi}^q}{\beta_\Pi} (\delta n^{\mu}_q \lambda_\mu) (\delta u^{\mu} \lambda_\mu)
\label{eq:unPi}
\end{align}
as well as the couplings with shear stress fluctuations 
\begin{align}
    e_{\pi^{\mu\nu},u^{\mu},n^{\mu}}&=  \frac{\beta_\pi}{2} \delta \pi_{\mu\nu} \delta \pi^{\mu\nu} \pm \delta\pi^{\mu\nu}\, \delta u_{\nu} \lambda_\mu  + \sum_q \delta_{n\pi}^q \lambda_\mu \delta n_\nu \delta \pi^{\mu\nu}\,, \nonumber\\
&= \frac{\beta_\pi}{2} \left[ \delta \pi^{\mu\nu} + \frac{1}{\beta_\pi} \lambda^\mu \delta u^{\nu} - \frac{\delta_{n\pi}^q}{\beta_\pi} \lambda^\mu \delta n^{\nu}_q \right ]^2 - \frac{1}{\beta_\pi} \lambda^2 \delta u^{\nu}\delta u_{\nu} \pm \frac{(\delta_{n\pi}^q)^2}{\beta_\pi} \lambda^2 \delta n^{\nu}_q \delta n_{\nu}^q- \frac{\delta_{n\pi}^q}{\beta_\pi} \lambda^2 \delta n_{\nu}^q \delta u^{\nu}\,.
\label{eq:unpi}
\end{align}
Inserting Eqs. (\ref{eq:unp}-\ref{eq:unpi}) back into the general form Eq (\ref{eq:Lyabunov-density}) then re-writing the functional by grouping relevant terms give us
\begin{align}
e&={\frac{1}{\varepsilon + p}  \left.\frac{\partial \varepsilon}{\partial p}\right\vert_{s} \left[\delta p + \frac{\varepsilon + p}{2} \left.\frac{\partial p}{\partial \varepsilon}\right\vert_{s}\left( \delta u^\mu \lambda_\mu + \left.\frac{\partial\alpha_q}{\partial p}\right\vert_s\delta n_q^\mu \lambda_\mu\right)\right]^2} \nonumber\\
&+{\frac{1}{\varepsilon + p} \left.\frac{\partial \varepsilon}{\partial s}\right\vert_{p} \left.\frac{\partial p}{\partial s}\right\vert_{\alpha_q} \left[ \delta s +  T (\varepsilon+p) \left.\frac{\partial s}{\partial \varepsilon}\right\vert_{p} \left.\frac{\partial s}{\partial p}\right\vert_{\alpha_q} \frac{\partial {\theta_q}}{\partial s}|_{p} (\delta n_q^\mu \lambda_\mu) \right]^2}\nonumber\\
&- { \frac{\varepsilon + p}{4} \left.\frac{\partial p}{\partial \varepsilon}\right\vert_{s}(\delta u^\mu \lambda_\mu)^2}-{\frac{1}{\beta_\Pi} (\delta u^{\mu} \lambda_\mu)^2}\nonumber\\
& -{ \frac{\varepsilon + p}{4} \left.\frac{\partial p}{\partial \varepsilon}\right\vert_{s} \Bigg(\left.\frac{\partial {\alpha_q}}{\partial p}\right\vert_{s}\Bigg)^2 (\delta n_q^\mu \lambda_\mu)^2}- { T^2 (\varepsilon+p) \left.\frac{\partial s}{\partial \varepsilon}\right\vert_{p} \left.\frac{\partial s}{\partial p}\right\vert_{\alpha_q} \Bigg(\left.\frac{\partial {\alpha_q}}{\partial s}\right\vert_{p}\Bigg)^2 (\delta n_q^\mu \lambda_\mu)^2}\pm {\frac{(\delta_{n\Pi}^q)^2}{\beta_\Pi} (\delta n^{\mu}_q \lambda_\mu)^2} \nonumber\\
&- {\frac{\varepsilon + p}{2} \left.\frac{\partial p}{\partial \varepsilon}\right\vert_{s} \left.\frac{\partial {\alpha_q}}{\partial p}\right\vert_{s} (\delta u^\mu \lambda_\mu)(\delta n_q^\mu \lambda_\mu)}- { \frac{\delta_{n\Pi}^q}{\beta_\Pi} (\delta n^{\mu}_q \lambda_\mu) (\delta u^{\mu} \lambda_\mu)}\nonumber\\
&+ {\frac{\beta_\pi}{2} \left[ \delta \pi^{\mu\nu} + \frac{1}{\beta_\pi} \lambda^\mu \delta u^{\nu} - \frac{\delta_{n\pi}^q}{\beta_\pi} \lambda^\mu \delta n^{\nu}_q \right ]^2}- {\frac{1}{\beta_\pi} \lambda^2 \delta u^{\nu}\delta u_{\nu} }- {\frac{(\delta_{n\pi}^q)^2}{\beta_\pi} \lambda^2 \delta n^{\nu}_q \delta n_{\nu}^q}- {\frac{\delta_{n\pi}^q}{\beta_\pi} \lambda^2 \delta n_{\nu}^q \delta u^{\nu}}\nonumber\\
&+{\frac{\beta_\Pi}{2} \left[ \delta \Pi + \frac{1}{\beta_\Pi} \delta u^{\mu} \lambda_\mu - \frac{\delta_{n\Pi}^q}{\beta_\Pi} \delta n^{\mu}_q \lambda_\mu\right ]^2 } \nonumber\\
&- (\varepsilon+ p) \frac{\delta u_{\nu} \delta u^{\nu}}{2 }+ \sum_q \frac{\beta_{n}^{qq'}}{2}  \delta n_q^\mu \delta {n}_\mu^{q'}
\label{eq:e-halfway}
\end{align}
As we can see from Eq. (\ref{eq:e-halfway}), in order to proceed further we need to project out time like and space like components of the flow velocity and charge currents. To do so, I will now manipulate the following relation
\begin{align}
\gamma^\mu_\nu&= g^\mu_\nu + u^{\mu} u_{\nu} - \frac{\lambda^{\mu} \lambda_{\nu}}{\lambda^2}\,,\nonumber\\
&= \Delta^\mu_\nu - \frac{\lambda^{\mu} \lambda_{\nu}}{\lambda^2}\,,\nonumber
\end{align}
\begin{align}
\gamma_\nu^\mu \delta u_{\mu} \delta u^{\nu} &= \Delta^\mu_\nu -  \frac{\lambda^{\mu} \lambda_{\nu}}{\lambda^2} ( \delta u_{\mu} \delta u^{\nu})\nonumber\\
& = g^\mu_\nu \delta u_{\mu} \delta u^{\nu} + u^{\mu} u_{\nu} \delta u_{\mu} \delta u^{\nu}-  \frac{\lambda^{\mu} \lambda_{\nu}}{\lambda^2} ( \delta u_{\mu} \delta u^{\nu})\nonumber\\
&=  \delta u_{\nu} \delta u^{\nu} + ( u_{\alpha} \delta u^{\alpha})^2- \frac{\lambda^{\mu}\delta u_{\mu} }{\lambda^2} 
\end{align}
Using the relation $u^\mu \delta u_\mu = - \delta u_\mu \delta u^\mu /2$ we cancel the middle term to get
\begin{align}
- \frac{\varepsilon+ p}{2} \delta u_{\nu} \delta u^{\nu}&= - \frac{\varepsilon+ p}{2} \gamma_\nu^\mu \delta u_{\mu} \delta u^{\nu} - \frac{\varepsilon+ p}{2}  \frac{(\lambda_\mu \delta u^{\mu})^2}{\lambda^2} 
\end{align}
Likewise
\begin{align}
\sum_q \frac{\beta_{n}^{qq'}}{2}  \delta n_q^\mu \delta {n}_\mu^{q'}&= \sum_q \frac{\beta_{n}^{qq'}}{2} \gamma_\nu^\mu \delta n^q_\mu \delta {n}^\nu_{q'} + \sum_q \frac{\beta_{n}^{qq'}}{2}  \frac{(\lambda_\mu \delta n_q^\mu)^2}{\lambda^2} \nonumber\\
\end{align}
And finally, once more, I  manipulate the relation
\begin{equation}
\gamma_{\mu\nu}= g_{\mu\nu} + u_{\mu} u_{\nu} - \frac{\lambda_{\mu} \lambda_{\nu}}{\lambda^2}\,,    
\end{equation}
to get
\begin{align}
\frac{(\delta_{n\pi}^q)^2}{\beta_\pi} \lambda^2 \delta n^{\nu}_q \delta n_{\nu}^q&= - \frac{(\delta_{n\pi}^q)^2}{\beta_\pi} \lambda^2 \gamma_\nu^\mu \delta n_{\mu}^q \delta n^{\nu}_{q'} - \frac{(\delta_{n\pi}^q)^2}{\beta_\pi} (\lambda_\mu \delta n^{\mu}_q)^2
\end{align}
\begin{align}
- \frac{\lambda^2}{\beta_\pi} \delta u_{\nu} \delta u^{\nu}&= - \frac{\lambda^2}{\beta_\pi} \gamma_\nu^\mu \delta u_{\mu} \delta u^{\nu} - \frac{1}{\beta_\pi}  (\lambda_\mu \delta u^{\mu})^2 
\end{align}
\begin{align}
\frac{\delta_{n\pi}^q}{\beta_\pi} \lambda^2 \delta n^{\nu}_q \delta u_{\nu}&= - \frac{\delta_{n\pi}^q}{\beta_\pi} \lambda^2 \gamma_\nu^\mu \delta n^{\nu}_q \delta u_{\nu} - \frac{\delta_{n\pi}^q}{\beta_\pi} (\lambda_\mu \delta n^{\mu}_q)(\lambda_\mu \delta u^{\mu}) 
\end{align}
Inserting the above expressions into the original form
\begin{align}
e&=\frac{1}{\varepsilon + p}  \left.\frac{\partial \varepsilon}{\partial p}\right\vert_{s} \left[\delta p + \frac{\varepsilon + p}{2} \left. \frac{\partial p}{\partial \varepsilon}\right\vert_{s}\left( \delta u^\mu \lambda_\mu + \left. \frac{\partial\alpha_q}{\partial p}\right\vert_s \delta n_q^\mu \lambda_\mu\right)\right]^2 \nonumber\\
&+\frac{1}{\varepsilon + p} \left.\frac{\partial \varepsilon}{\partial s}\right\vert_{p} \left.\frac{\partial p}{\partial s}\right\vert_{\alpha_q} \left[ \delta s +  T (\varepsilon+p) \left.\frac{\partial s}{\partial \varepsilon}\right\vert_{p} \left.\frac{\partial s}{\partial p}\right\vert_{\alpha_q} \left.\frac{\partial {\alpha_q}}{\partial s}\right\vert_{p} (\delta n_q^\mu \lambda_\mu) \right]^2\nonumber\\
&-  \frac{\varepsilon + p}{4} \left. \frac{\partial p}{\partial \varepsilon}\right\vert_{s}(\delta u^\mu \lambda_\mu)^2-{\frac{1}{\beta_\Pi} (\delta u^{\mu} \lambda_\mu)^2}- \frac{\varepsilon+ p}{2}  \frac{(\lambda_\mu \delta u^{\mu})^2}{\lambda^2} -{ \frac{\varepsilon + p}{4} \left.\frac{\partial p}{\partial \varepsilon}\right\vert_{s} \Bigg(\left.\frac{\partial {\alpha_q}}{\partial p}\right\vert_{s}\Bigg)^2 (\delta n_q^\mu \lambda_\mu)^2}\nonumber\\
&- { T^2 (\varepsilon+p) \left.\frac{\partial s}{\partial \varepsilon}\right\vert_{p} \left.\frac{\partial s}{\partial p}\right\vert_{\alpha_q} \Bigg(\left.\frac{\partial {\alpha_q}}{\partial s}\right\vert_{p}\Bigg)^2 (\delta n_q^\mu \lambda_\mu)^2}\pm {\frac{(\delta_{n\Pi}^q)^2}{\beta_\Pi} (\delta n^{\mu}_q \lambda_\mu)^2} - \frac{(\delta_{n\pi}^q)^2}{\beta_\pi} (\lambda_\mu \delta n^{\mu}_q)^2+ \sum_q \frac{\beta_{n}^{qq'}}{2}  \frac{(\lambda_\mu \delta n_q^\mu)^2}{\lambda^2}\nonumber\\
&- {\frac{\varepsilon + p}{2} \left.\frac{\partial p}{\partial \varepsilon}\right\vert_{s} \left.\frac{\partial {\alpha_q}}{\partial p}\right\vert_{s} (\delta u^\mu \lambda_\mu)(\delta n_q^\mu \lambda_\mu)}- { \frac{\delta_{n\Pi}^q}{\beta_\Pi} (\delta n^{\mu}_q \lambda_\mu) (\delta u^{\mu} \lambda_\mu)}- \frac{\delta_{n\pi}^q}{\beta_\pi} (\lambda_\mu \delta n^{\mu}_q)(\lambda_\mu \delta u^{\mu})\nonumber\\
&+ {\frac{\beta_\pi}{2} \left[ \delta \pi^{\mu\nu} + \frac{1}{\beta_\pi} \lambda^\mu \delta u^{\nu} \pm \frac{\gamma_{n\pi}^q}{\beta_\pi} \lambda^\mu \delta n^{\nu}_q \right ]^2}+{\frac{\beta_\Pi}{2} \left[ \delta \Pi + \frac{1}{\beta_\Pi} \delta u^{\mu} \lambda_\mu \pm \frac{\delta_{n\Pi}^q}{\beta_\Pi} \delta n^{\mu}_q \lambda_\mu\right ]^2 } - \frac{\varepsilon+ p}{2} \gamma_\nu^\mu \delta u_{\mu} \delta u^{\nu}\nonumber\\
&+\sum_q \frac{\beta_{n}^{qq'}}{2} \gamma_\nu^\mu \delta n^q_\mu \delta {n}^\nu_{q'} - {\frac{\lambda^2}{\beta_\pi} \gamma_\nu^\mu \delta u_{\mu} \delta u^{\nu} - \frac{1}{\beta_\pi}  (\lambda_\mu \delta u^{\mu})^2 } { - \frac{(\delta_{n\pi}^q)^2}{\beta_\pi} \lambda^2 \gamma_\nu^\mu \delta n_{\mu}^q \delta n^{\nu}_{q'}  }- {\frac{\delta_{n\pi}^q}{\beta_\pi} \lambda^2 \gamma_\nu^\mu \delta n^{\nu}_q \delta u_{\nu} }
\label{eq:semi}
\end{align}
Now by looking at the terms 
\begin{align}
e&=\bigg(\sum_q \frac{\beta_{n}^{qq'}}{2} { - \frac{(\delta_{n\pi}^q)^2}{\beta_\pi} \lambda^2\bigg) \gamma_\nu^\mu \delta n_{\mu}^q \delta n^{\nu}_{q'}}- {\frac{\delta_{n\pi}^q}{\beta_\pi} \lambda^2 \gamma_\nu^\mu \delta n^{\nu}_q \delta u_{\mu}} \nonumber\\
&= \bigg(\sum_q \frac{2\beta_\pi\beta_{n}^{qq'}-(\delta_{n\pi}^q)^2\lambda^2}{2\beta_\pi}\bigg)\gamma_\nu^\mu \delta n_{\mu}^q \delta n^{\nu}_{q'}-\frac{\delta_{n\pi}^q}{\beta_\pi} \lambda^2 \gamma_\nu^\mu \delta n^{\nu}_q \delta u_{\mu} \nonumber\\
&= \bigg(\sum_q \frac{2\beta_\pi\beta_{n}^{qq'}-(\delta_{n\pi}^q)^2\lambda^2}{2\beta_\pi}\bigg) \bigg[\gamma_\nu^\mu \delta n_{\mu}^q \delta n^{\nu}_{q'}-\frac{\delta_{n\pi}^q \lambda^2}{2\beta_\pi\beta_{n}^{qq'}-(\delta_{n\pi}^q)^2\lambda^2} \gamma_\nu^\mu \delta n^{\nu}_q \delta u_{\mu}\bigg]  
\end{align}
where completing the square the gives
\begin{align}
e&=  \bigg[\gamma_\nu^\mu \delta n_{\mu}^q -\frac{\delta_{n\pi}^q \lambda^2}{2\beta_\pi\beta_{n}^{qq'}-(\delta_{n\pi}^q)^2\lambda^2} \gamma_\nu^\mu \delta u_{\mu}\bigg] \bigg[\gamma^\nu_\mu \delta n^{\mu}_q -\frac{\delta_{n\pi}^q \lambda^2}{2\beta_\pi\beta_{n}^{qq'}-(\delta_{n\pi}^q)^2\lambda^2} \gamma^\nu_\mu \delta u^{\mu}\bigg]\nonumber\\
&= \gamma_\nu^\mu \gamma^\nu_\mu \delta n_{\mu}^q  \delta n^{\mu}_{q'} + \bigg(\frac{\delta_{n\pi}^q \lambda^2}{2\beta_\pi\beta_{n}^{qq'}-(\delta_{n\pi}^q)^2\lambda^2}\bigg)^2 \gamma_\nu^\mu \delta u_{\mu}\gamma^\nu_\mu \delta u^{\mu}- 2 \bigg(\frac{\delta_{n\pi}^q \lambda^2}{2\beta_\pi\beta_{n}^{qq'}-(\delta_{n\pi}^q)^2\lambda^2}\bigg) \gamma_\nu^\mu \delta u_{\mu}\gamma^\nu_\mu \delta u^{\mu}\nonumber\\
&*= \bigg(\sum_q \frac{2\beta_\pi\beta_{n}^{qq'}-(\delta_{n\pi}^q)^2\lambda^2}{2\beta_\pi}\bigg)\bigg[\gamma_\nu^\mu \gamma^\nu_\mu \delta n_{\mu}^q  \delta n^{\mu}_{q'} + \bigg(\frac{\delta_{n\pi}^q \lambda^2}{2\beta_\pi\beta_{n}^{qq'}-(\delta_{n\pi}^q)^2\lambda^2}\bigg)^2 \gamma_\nu^\mu \delta u_{\mu}\gamma^\nu_\mu \delta u^{\mu}\nonumber\\
&- 2 \bigg(\frac{\delta_{n\pi}^q \lambda^2}{2\beta_\pi\beta_{n}^{qq'}-(\delta_{n\pi}^q)^2\lambda^2}\bigg) \gamma_\nu^\mu \delta u_{\mu}\gamma^\nu_\mu \delta n_q^{\mu}\bigg]\nonumber\\
&*=\bigg(\sum_q \frac{2\beta_\pi\beta_{n}^{qq'}-(\delta_{n\pi}^q)^2\lambda^2}{2\beta_\pi}\bigg)\bigg[\gamma_\nu^\mu \gamma^\nu_\mu \delta n_{\mu}^q  \delta n^{\mu}_{q'}\bigg] +\bigg( \frac{(\delta_{n\pi}^q \lambda^2)^2}{2 \beta_\pi(2\beta_\pi\beta_{n}^{qq'}-(\delta_{n\pi}^q)^2\lambda^2)}\bigg) \bigg[\gamma_\nu^\mu \delta u_{\mu}\gamma^\nu_\mu \delta u^{\mu}\bigg]\nonumber\\
&- 2 \bigg(\frac{\delta_{n\pi}^q \lambda^2}{2\beta_\pi}\bigg) \gamma_\nu^\mu \delta u_{\mu}\gamma^\nu_\mu \delta n_q^{\mu}\bigg]
\end{align}
We then arrive at
\begin{align}
\bigg(\sum_q \frac{\beta_{n}^{qq'}}{2} { - \frac{(\delta_{n\pi}^q)^2}{\beta_\pi} \lambda^2\bigg) \gamma_\nu^\mu \delta n_{\mu}^q \delta n^{\nu}_{q'}}- {\frac{\delta_{n\pi}^q}{\beta_\pi} \lambda^2 \gamma_\nu^\mu \delta n^{\nu}_q \delta u_{\mu}}&= \bigg(\sum_q \frac{2\beta_\pi\beta_{n}^{qq'}-(\delta_{n\pi}^q)^2\lambda^2}{2\beta_\pi}\bigg)\nonumber\\
&\times \bigg[\gamma_\nu^\mu \delta n_{\mu}^q -\frac{\delta_{n\pi}^q \lambda^2}{2\beta_\pi\beta_{n}^{qq'}-(\delta_{n\pi}^q)^2\lambda^2} \gamma_\nu^\mu \delta u_{\mu}\bigg] ^2\nonumber\\
&- \bigg( \frac{(\delta_{n\pi}^q \lambda^2)^2}{2 \beta_\pi(2\beta_\pi\beta_{n}^{qq'}-(\delta_{n\pi}^q)^2\lambda^2)}\bigg) \bigg[\gamma_\nu^\mu \delta u_{\mu}\gamma^\nu_\mu \delta u^{\mu}\bigg] ,    
\end{align}
which provides two additional constraints.
Likewise, I collect the terms
\begin{align}
e&= \bigg(\frac{\beta_{n}^{qq'}}{2\lambda^2} - \frac{({\delta_{n\pi}^q})^2}{\beta_\pi}-  \frac{(\delta_{n\Pi}^q)^2}{\beta_\Pi} -{ \frac{\varepsilon + p}{4} \left.\frac{\partial p}{\partial \varepsilon}\right\vert_{s} \Bigg(\left.\frac{\partial {\alpha_q}}{\partial p}\right\vert_{s}\Bigg)^2 }- T^2 (\varepsilon+p) \left.\frac{\partial s}{\partial \varepsilon}\right\vert_{p} \left.\frac{\partial s}{\partial p}\right\vert_{\alpha_q} \Bigg(\left.\frac{\partial {\alpha_q}}{\partial s}\right\vert_{p}\Bigg)^2\bigg)(\delta n^{\mu}_q \lambda_\mu)^2\nonumber\\
&- \bigg(\frac{\delta_{n\Pi}^q}{\beta_\Pi}+\frac{\delta_{n\pi}^q}{\beta_\pi}-  \frac{\varepsilon + p}{2} \left.\frac{\partial p}{\partial \varepsilon}\right\vert_{s} \left.\frac{\partial {\alpha_q}}{\partial p}\right\vert_{s} \bigg)(\delta n^{\mu}_q \lambda_\mu) (\delta u^{\mu} \lambda_\mu).\nonumber
\end{align}
To make things a bit more compact, I will define
\begin{align}
e_{n,u}&= \frac{\beta_{n}^{qq'}}{2\lambda^2} - \frac{({\delta_{n\pi}^q})^2}{\beta_\pi}-  \frac{(\delta_{n\Pi}^q)^2}{\beta_\Pi} -{ \frac{\varepsilon + p}{4} \left.\frac{\partial p}{\partial \varepsilon}\right\vert_{s} \Bigg(\left.\frac{\partial {\alpha_q}}{\partial p}\right\vert_{s}\Bigg)^2 }- T^2 (\varepsilon+p) \left.\frac{\partial s}{\partial \varepsilon}\right\vert_{p} \left.\frac{\partial s}{\partial p}\right\vert_{\alpha_q} \Bigg(\left.\frac{\partial {\alpha_q}}{\partial s}\right\vert_{p}\Bigg)^2
\end{align}
\begin{align}
C_{n,u}&=  - \frac{\delta_{n\Pi}^q}{\beta_\Pi}+\frac{\delta_{n\pi}^q}{\beta_\pi}-\frac{\varepsilon + p}{2} \left.\frac{\partial p}{\partial \varepsilon}\right\vert_{s} \left.\frac{\partial {\alpha_q}}{\partial p}\right\vert_{s} \nonumber
\end{align}
\begin{align}
e&= e_{n,u} \,\big[ \delta n_{\mu}^q \delta n^{\mu}_{q'}- \frac{C_{n,u}}{e_{n,u}}  \delta n^{\mu}_q \delta u_{\mu}\big]\nonumber
\end{align}
\begin{align}
\bigg( \delta n_{\mu}^q - \frac{C_{n,u}}{e_{n,u}}  \delta u_{\mu}\bigg)^2= \delta n_{\mu}^q \delta n^{\mu}_{q'} + \bigg( \frac{C_{n,u}}{e_{n,u}}\bigg)^2 \delta u^{\mu} \delta u_{\mu} - 2\,  \frac{C_{n,u}}{e_{n,u}}\delta n^{\mu}_q \delta u_{\mu} \nonumber
\end{align}
to get to
\begin{align}
\delta n_{\mu}^q \delta n^{\mu}_{q'}- 2\,  \frac{C_{n,u}}{e_{n,u}}\delta n^{\mu}_q \delta u_{\mu} = \bigg( \delta n_{\mu}^q - \frac{C_{n,u}}{e_{n,u}} \delta u_{\mu}\bigg)^2- \bigg( \frac{C_{n,u}}{e_{n,u}}\bigg)^2 \delta u^{\mu} \delta u_{\mu}.
\label{eq:deltau}
\end{align}
Finally, I group all terms with $\delta u^\mu \lambda_\mu$ from Eq. (\ref{eq:semi}) and the expression in Eq. (\ref{eq:deltau}) to obtain the final condition
\begin{align}
- { \frac{\varepsilon + p}{4} \left.\frac{\partial p}{\partial \varepsilon}\right\vert_{s}(\delta u^\mu \lambda_\mu)^2}-{\frac{1}{\beta_\Pi} (\delta u^{\mu} \lambda_\mu)^2}-{\frac{2}{3\beta_\pi} (\delta u^{\mu} \lambda_\mu)^2}- \frac{\varepsilon+ p}{2}  \frac{(\lambda_\mu \delta u^{\mu})^2}{\lambda^2} -  \frac{C_{n,u}^2}{e_{n,u}} \delta u^{\mu} \delta u_{\mu}
\end{align}
Which provides a constraint associated with $\delta u^\mu \lambda_\mu$.

\end{document}